\documentclass[aps,pre,onecolumn,showpacs,showkeys,a4paper]{revtex4}
\input epsf  
\usepackage{graphicx} 
\usepackage{amsmath}
\usepackage{amssymb}
\usepackage{amscd}

\begin{document} 

\title{Fluctuations in Polymer Translocation}
\author{P. L. Krapivsky$^{1}$ and K. Mallick$^{2}$}
\affiliation{$^1$ Department of Physics, Boston University,  
Boston, MA 02215, USA\\
$^{2}$Institut de Physique Th\'eorique CEA, IPhT, F-91191 Gif-sur-Yvette,
 France}
\begin{abstract}
We investigate a model of chaperone-assisted polymer translocation through a nanopore in a membrane. Translocation is driven by irreversible random sequential absorption of chaperone proteins that bind to the polymer on one side of the membrane. The proteins are larger than the pore and hence the backward motion of the polymer is inhibited. This mechanism rectifies Brownian fluctuations and results in an effective force that drags the polymer in a preferred direction. The translocated polymer undergoes an effective biased random walk and we compute the corresponding diffusion constant. Our methods allow us to determine the large deviation function which, in addition to velocity and diffusion constant, contains the entire statistics of the translocated length.  
\end{abstract} 
   \pacs{05.60.--k, 87.16.Ac, 05.10.Ln}
  \keywords{Translocation, Ratchet Effect, Non-Equilibrium Fluctuations}

\maketitle
  
Proteins, nucleic acids, and various  products synthesized inside cells are  transported within the cytoplasm by molecular machines or motors \cite{motors}. Polymeric chains may also translocate in order to get in or out of organelles within the cell, or to cross the cell's outer membrane \cite{Alberts}. 
Both transport and translocation cannot rely on diffusion alone: 
macromolecules are subject to thermal fluctuations which are isotropic and from which, according to the second law of thermodynamics, no useful work can be extracted. Pure diffusion can account neither for the directionality of motion nor for the observed time scales --- supplementary  mechanisms that induce
active and  directed motion must be taken into account. The `Brownian Ratchet'   \cite{feynman}  provides a general setting to describe  rectification  of diffusive motion using chemical energy. This paradigm helps to understand the physics of molecular motors and to construct mathematical models of the translocation process. As explained e.g. in Refs.~\cite{PeskinPNAS,siam},
Brownian motion will cause a protein undergoing the translocation through 
the pore to fluctuate back and forth, and chemical asymmetries will rectify its  displacement. Several mechanisms  \cite{PNelson} can play this role including, 
for instance, disulfide bond formation, attachment of sugar bonds, binding of 
chaperone proteins on one side of the membrane. These rectification
processes induce an effective force on the polymeric chain that
drags it in the desired direction by inhibiting backwards motion.

The chaperone-assisted translocation 
\cite{PeskinPNAS,siam,nelson,elston,rap,rudnick,metzler,chou} is a common mechanism for protein translocation and perhaps for DNA transport through membranes \cite{DNA-transl}.  Earlier work \cite{PeskinPNAS,siam,nelson,rudnick} has focused on continuous space and time descriptions. This  assumption effectively means that the chaperone molecule is much longer than the lattice size. In the case of nucleic acids, for instance, the lattice size is the interbase distance $\sim 0.36$ nm which is about 6 times shorter than the chaperone molecule. Hence the discreteness of the monomers which are the binding sites for the chaperones  can have sizable effect on the kinetics of the process. D'Orsogna, Chou, and Antal \cite{chou} have recently investigated a discrete model of the polymer translocation driven by the irreversible random sequential absorption of chaperone proteins on one side of the pore. The chief analytical result of Ref.~\cite{chou} is the computation of the translocation velocity. Needless to say, the translocation velocity (here,  the average translocated length per unit time) provides a partial description which is not sufficient in many biologically relevant cases, especially when the translocated polymers are not too long.  In these situations dispersion (quantified by diffusion coefficients) and more generally large deviations play an important role. The goal of the present work is to carry out a thorough mathematical analysis of this polymer translocation model. We shall focus on the diffusion coefficient, although our approach gives the full statistics of the translocated length, for instance all its   higher cumulants. 

The outline of this work is as follows. In  section~\ref{sec:general},
we describe the  translocation model and discuss some of its general scaling properties.  In section~\ref{sec:discrete}, we investigate the simplest
case when chaperone molecules attach to single monomers along the polymer chain. First, we  calculate the velocity as a function of the parameters of the model. Then we determine the large deviation function which encodes 
the full statistics of the translocated length; in particular, the large deviation function contains the velocity and the diffusion constant.  In section~\ref{sec:continuous},
we study a continuous space model where the chaperones are free to bind
anywhere on the translocated segment of the polymer. This corresponds to the limit 
when the size of the chaperone molecule is large with respect to the monomer size
(providing the natural minimal size in the problem, the lattice spacing). In this continuum case the velocity and the fluctuations can also be  determined analytically. We show that the predictions of the discrete and the continuum 
frameworks match when the chaperon attachment rate $\lambda$ is vanishingly small. In the complimentary limit of large attachment rate, the system becomes equivalent to a continuous time random walk. This limit admits a separate more elementary analysis and it provides a useful check of the consistency of our calculations. Finally in section~\ref{sec:discussion} we discuss possible extensions.
A generalization to the situation when the polymer undergoes a bias diffusion and various technical calculations are relegated to Appendices.

 \section{General description of the system}
  \label{sec:general}

\begin{figure}[ht]
\begin{center}
\includegraphics[width=0.75\textwidth]{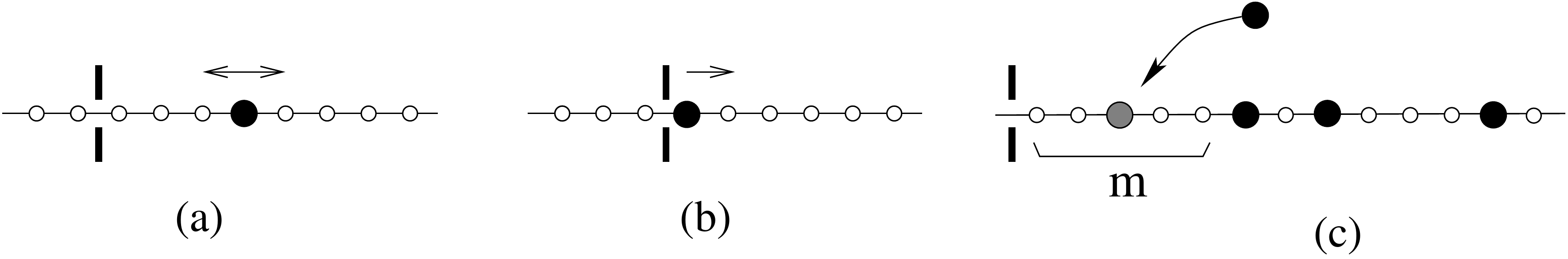}
\caption{Illustration of chaperon-assisted translocation.  (a) The polymer
  can hop in either direction.  (b) The polymer can hop only to the right
  because an adsorbed chaperone (large dot) is next to the pore (hole  in the
  membrane barrier) and is too large to enter. (c) Adsorption of a new chaperone (shaded)
  at  a site on   the leftmost chaperone-free segment.}
\label{fig-ads-transloc}
\end{center}
\end{figure}

Consider a polymer chain that passes through a pore in a membrane
(see figure~\ref{fig-ads-transloc}). We assume that the motion of the  polymer chain is equivalent to unbiased random walk, i.e., it is fully described by one parameter, the diffusion coefficient $D$. The pore is located  at $x=0$ and we focus on the  polymer segment to the right of the membrane (the region located on the 
right-hand side of the pore is the {\it target region} of the translocation process). At a  given moment, this polymer segment consists of a certain number $L$ of monomer units, each of size $a$,   
labeled $1,\ldots,L$. Translocation is reversible and  the polymer can move by one-monomer unit to the right or left with equal rates, resulting  in a zero average translocation velocity and a non-vanishing  diffusion coefficient $D$ for  the `bare' polymer. We now suppose that the medium on the right side of the membrane contains a fixed density of special molecules --- ``chaperones'' --- that adsorb irreversibly, with a certain rate $\lambda$, onto unoccupied adjacent sites
of the polymer.  A chaperone is sufficiently large that
it cannot pass through the pore (Fig.~\ref{fig-ads-transloc}). 
As a result, the chaperones rectify the
polymer diffusion so that it passes through the pore at a non-zero speed $v$
and has a certain effective  diffusion coefficient $\mathcal{D}$. 
More precisely,  $L$,  the number of translocated monomers at time $t$, is a random variable. Let $P(L,t)$ be the corresponding probability distribution. One anticipates that this distribution is asymptotically Gaussian,
\begin{equation}
\label{Gaussian}
P(L,t) \to \frac{1}{\sqrt{4\pi \mathcal{D} t}}\,e^{-(L-vt)^2/4\mathcal{D}t}\,,
\end{equation}
where the velocity and the diffusion coefficient of the tip characterize, respectively, 
the average and the variance of $L$ 
\begin{equation}
\label{L-av-var}
\langle L\rangle= v t,\quad \langle L^2\rangle-\langle L\rangle^2=2\mathcal{D} t\,.
\end{equation}
The  quantities $v$ and $\mathcal{D}$  depend  on $\lambda$,
the length $\ell$ of the chaperone molecule, the monomer length  $a$, and the bare diffusion coefficient $D$. On dimensional grounds, we can write 
\begin{equation}
\label{dim}
v = \tfrac{D}{\ell}\,F\big(\bar{\lambda}, \tfrac{\ell}{a}\big),   \quad 
\mathcal{D} = D\, G\big(\bar{\lambda},\tfrac{\ell}{a}\big) \quad
 \hbox{ with } \quad \bar{\lambda} = \frac{\lambda \ell^3}{D} \, . 
\end{equation}
These  simple relations  predict scaling behaviors in the interesting regime of 
small adsorption rate, that is, when $\bar{\lambda} \ll 1$. In this limit we
anticipate that the length of the chaperone molecules should become
irrelevant. This implies that $F\sim \bar{\lambda}^{1/3}$ as $\bar{\lambda}  \to 0$, while $G$ remains finite. Therefore
\begin{equation}
\label{scaling}
v \sim \lambda^{1/3} D^{2/3}  \quad \hbox{ and }
    \quad \mathcal{D} \sim D \quad {\rm when}\quad 
 \bar{\lambda} = \frac{\lambda \ell^3}{D} \to 0\, . 
\end{equation}
In the opposite limit of large adsorption rate, $\bar{\lambda}  \gg 1$, we anticipate that adsorption becomes irrelevant. This suggests that
\begin{equation}
\label{scaling2}
v \sim \frac{D}{\ell}\,  \quad \hbox{ and }   \quad \mathcal{D} \sim D \quad {\rm when}\quad 
\bar{\lambda}  = \frac{\lambda \ell^3}{D}\to \infty  \, .
\end{equation}

In the following sections we derive exact expressions for the functions $F(\bar{\lambda})$ and $G(\bar{\lambda})$ defined in Eq.~(\ref{dim}). We will confirm 
the scaling behaviors of \eqref{scaling}--\eqref{scaling2} and determine analytical expressions for the prefactors. 

\section{Analysis of the  discrete space model}
\label{sec:discrete}

In this section we analyze the simplest discrete model corresponding to the situation when chaperone molecules have the same length as the monomer units of the polymer chain ($\ell =a$). Bound molecules do not overlap, that is, there is at most one chaperone per site. Another assumption is that the binding of chaperone molecules to the polymer is irreversible. No other 
assumptions will be made and the following analysis will involve no approximations. 

The polymer hops to the left and right with equal rates which we set equal to unity; we further set $\ell=1$. With these agreements, the intrinsic diffusion constant $D$ of the bare polymer is also equal to 1. Chaperones attach 
to empty sites with rate $\lambda$ and we assume that attachment is
irreversible (the detachment rate is 0). Since we set $\ell=1$ and $D=1$, the adsorption rate $\lambda$ becomes dimensionless, $\lambda=\bar{\lambda}$. 

At any given moment, the complete description is
provided by the length $L$ of the  translocated segment of the polymer and the positions of the  occupied  monomers by adsorbed chaperones:
\begin{equation}
\label{positions}
1\leq m_1<m_2<\ldots <m_M\leq L \, .
\end{equation}
The length $L$  of the translocated segment, the number $M$ of the attached chaperone molecules (equivalently, the number of occupied monomers) and their positions $m_1,\ldots, m_M$ along the polymer chain are all random variables.  At a given time, the first monomer to the right of the pore is always labeled as 1. The above definition \eqref{positions} shows that an occupied 
monomer is not allowed  to move to  the position $x=0$; in other words, if $m_1=1$ the segment can hop only to the right, 
while for $m_1>1$ both hops are equally possible. 
The full description requires the computation of the probability density 
$P(L; m_1,\ldots,m_M;t)$. A shorter description ignores all monomers except the one closest to the pore, located at a certain 
site $j$ (i.e.  $m_1=j$). In other words, this shorter description focuses on the segment probability $Q_j(t)$, namely the probability that  at time $t$, the leftmost
chaperone is located at distance $j$ from the pore. Schematically,
\begin{equation*}
Q_j(t)={\rm Prob}\{|\underbrace{\circ\ldots\circ}_{j-1}\bullet\}
\end{equation*}
where the vertical line represents the position of the membrane pore. 
The key feature of this problem is that the time-evolution equations of  the $Q_j(t)$'s form a closed set and can be solved.  (For a review of techniques developed in studies of adsorption kinetics, see \cite{Evans,PaulK}.) These equations read
\begin{eqnarray}
  \frac{d Q_j(t)}{dt} &=&  Q_{j-1}(t)  + Q_{j+1}(t) + \lambda \sum_{k>j} Q_k(t)
  -[2+\lambda(j-1)] Q_j(t) 
  \, \hbox{ for } \,\,\, j>1, 
  \label{eq:Qj} \\
    \frac{d Q_1(t)}{dt} &=& Q_2(t) +  \lambda \sum_{k>1} Q_k(t) - Q_1(t) \, .
    \label{eq:Q1}
\end{eqnarray}
The terms in Eqs.~\eqref{eq:Qj} are self-explanatory, e.g. the first two terms 
on the right-hand side describe the gain due to hopping, while the third  gain 
term represents adsorption events. Equation \eqref{eq:Q1} does not fully fit into the pattern, but  if we add
an extra variable $Q_0(t)$ with $Q_0(t) = Q_1(t),$ then equation  \eqref{eq:Q1}
does  take the same form as Eqs.~\eqref{eq:Qj}.  As a useful check of self-consistency one can add Eqs.~\eqref{eq:Qj}--\eqref{eq:Q1} and verify that 
the sum $\sum_{k>0} Q_k(t)$ is conserved.
(By normalization, $\sum_{k>0} Q_k(t) = 1$ must hold for all values of $t$.)

It is useful to define the cumulative variable  $E_m(t) = \sum_{k>m} Q_k(t) $. 
Note that $Q_m(t) = E_{m-1}(t) - E_m(t)$. The  quantities  $E_m(t)$  represent 
empty interval probabilities
\begin{equation*}
E_m(t)={\rm Prob}\{|\underbrace{\circ\ldots\circ}_m\} \, . 
\end{equation*}
In particular, we have   $E_0(t) =  \sum_{k>0} Q_k(t)= 1$ and 
$E_{-1}(t) = Q_0(t) + E_0(t) = Q_0(t) +1$.
The equations of motion of   $E_m(t)$ are obtained from
Eqs.~\eqref{eq:Qj}--\eqref{eq:Q1} by summing these equations for ${k>m}$:
\begin{equation}
  \frac{d E_m(t)}{dt} =
  E_{m-1}(t) +  E_{m+1}(t) -\left(2+ \lambda m  \right) E_m(t)\,.
\label{eq:Emevol}
\end{equation}
Taking into account the boundary values, we see that equation \eqref{eq:Emevol}
is valid for all $m \ge 0$. The interpretation of the evolution equation \eqref{eq:Emevol} is simple: the gain terms arise from diffusion; the loss term describes diffusion and attachment (the latter occurs with rate $\lambda$ onto any of the $m$ sites of the empty interval). For a more detailed derivation of equations \eqref{eq:Qj}--\eqref{eq:Emevol} see Ref.~\cite{chou} and chapter 7 of Ref.~\cite{PaulK}.

\subsection{Steady state weights and velocity}
\label{vel}

\begin{figure}[ht]
\begin{center}
\begin{tabular}{cc}
\includegraphics[width=0.375\textwidth]{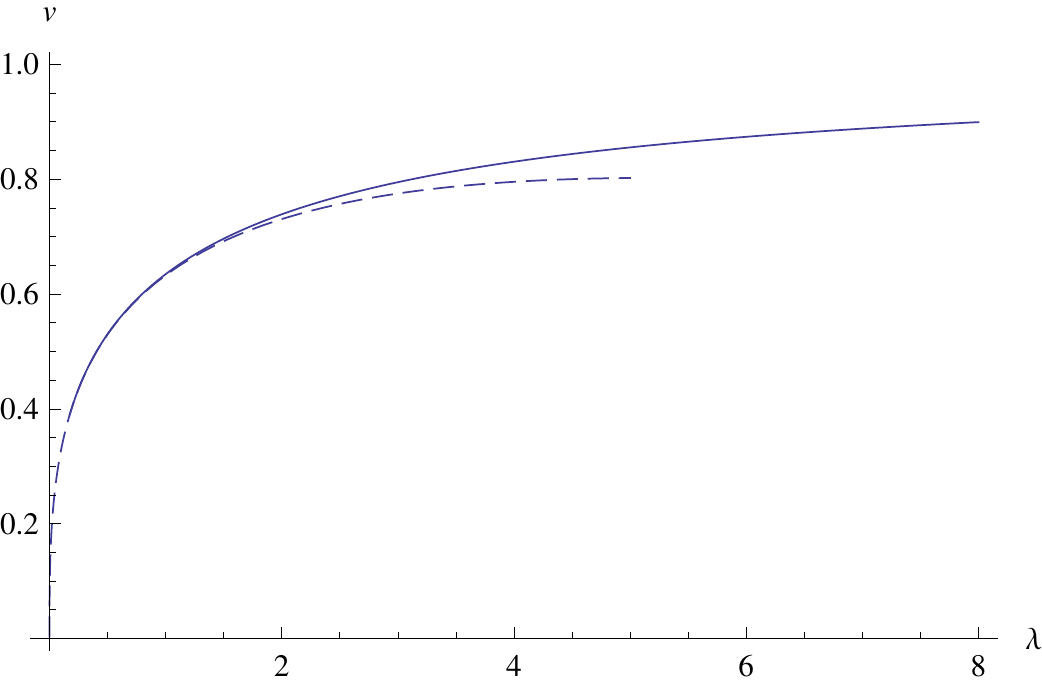} \quad \quad   &
  \quad  \quad 
\includegraphics[width=0.375\textwidth]{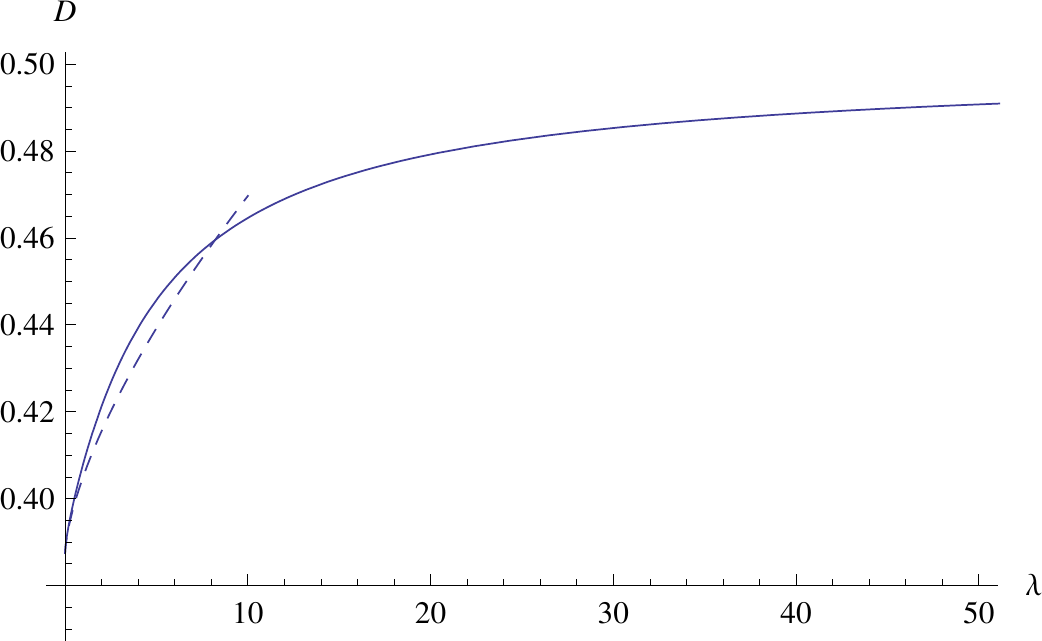} \\
\end{tabular}
\caption{Velocity and diffusion constant versus  the attachment
  rate. The left figure displays the exact formula~\eqref{eq:vitesse} 
for the velocity, the dashed line represents
  the asymptotic formula \eqref{V-smallbis}.   The right figure shows 
the diffusion
  constant~\eqref{formule:D}; the small $\lambda$ expansion (dashed 
line) given by Eq.~\eqref{D-small}
  is also plotted for comparison.}
\label{figuremath1}
\end{center}
\end{figure}

In the stationary state, the values of the  $E_m$'s must satisfy
 the following recursion:
\begin{equation}
    E_{m-1} +  E_{m+1} =  \left(2+ \lambda m  \right) E_m \, .
\label{eq:Em}
\end{equation}
We emphasize again that the boundary case  $m=0$ also fits into the pattern
and therefore we have $E_1+ E_{-1} = 2$.
Equation \eqref{eq:Em} is essentially the difference analog of an Airy equation.
The solution of this difference equation that satisfies the boundary condition  
$E_0 = 1$   can be expressed in terms of Bessel functions. This can be done by comparing \eqref{eq:Em} with the well-known 
identity \cite{abramovich}
\begin{equation}
 J_{\nu+1} (x) +  J_{\nu-1} (x)  = \frac{2\nu}{x} J_{\nu} (x) 
\label{Id:Bessel}
\end{equation}
for the Bessel functions. We notice that Eq.~(\ref{eq:Em}) is solved for all values of $m$ (including $m=0$) by choosing
\begin{equation}
    E_m = \frac{ J_{m+\Lambda}(\Lambda)}{ J_{\Lambda}(\Lambda)}
 \,\,\, \hbox{ with } \,\,\,  \Lambda =\frac{2}{\lambda} \,.
\label{eq:solEm}
\end{equation}
Having computed these steady state probabilities, we derive  the velocity of the polymer
\begin{equation}
 v = Q_1 = 1 -E_1 =
  1 - \frac{ J_{\Lambda+1}(\Lambda)}{ J_{\Lambda}(\Lambda)} \, . 
 \label{eq:vitesse}
\end{equation}
This result agrees with that found in Ref.~\cite{chou}. The expression  
\eqref{eq:vitesse} is compact but fairly complicated as the adsorption rate also enters into the index of the Bessel function. The plot of $v=v(\lambda)$, see
figure~\ref{figuremath1}, shows that in accord with intuition, the velocity is a monotonously increasing function of the adsorption rate. 

To better appreciate the behavior of the velocity, let us consider the limits of small and large adsorption rate. The small $\lambda$ behavior is derived from an asymptotic formula
 \cite{watson}
\begin{equation}
J_X(X+x)=\left(\frac{2}{X}\right)^{1/3} {\rm Ai}(0)- \left(\frac{2}{X}\right)^{2/3}
  {\rm Ai}'(0)x+\ldots
\label{Expans1}
\end{equation}
which is valid for  $X\to\infty$ and for  a fixed value of $x$. We  thus  obtain
\begin{equation}
\label{V-small}
v  = - \frac{{\rm Ai}'(0)}{{\rm Ai}(0)}\,\lambda^{1/3} \, .
\end{equation}
Using  the well-known expressions \cite{abramovich}
\begin{equation}
{\rm Ai}(0)=\frac{1}{3^{2/3}\Gamma(2/3)}\,,\quad 
{\rm Ai}'(0)=-\frac{1}{3^{1/3}\Gamma(1/3)} \, , 
\end{equation}
equation~\eqref{V-small} can be rewritten as 
\begin{equation}
\label{V-final}
v = {\mathcal A} \lambda^{1/3}\,  \quad  \hbox{ with } \qquad
 {\mathcal A}=\frac{3^{1/3} \Gamma(2/3)}{\Gamma(1/3)} = 
0.72901113\ldots
\end{equation}
The dependence of $v$ on $\lambda$ agrees with the scaling argument given in Eq.~(\ref{scaling}). This behavior can be understood more physically as follows.
Let $\tau$ be  the typical time between two successive  adsorptions on the leftmost empty segment. The tip will move a distance $d  \sim \sqrt{\tau}$. Now we note  that $\lambda d \tau\sim 1$ since $\lambda$ is the adsorption rate per site. These relations imply  that $\tau\sim \lambda^{-2/3}$ and $d \sim \lambda^{-1/3}$. Therefore
\begin{equation*}
  v \sim \frac{d}{\tau}\sim \lambda^{1/3}
\end{equation*}
explaining the $\lambda^{1/3}$ scaling. Higher orders of the expansion of the velocity with respect to the attachment rate $\lambda$
can be derived from the asymptotic expansion 
\begin{equation}
J_X(X+x)= \frac{1}{3\pi} \sum_{m\ge 0} B_m(x) \sin \left(\frac{m+1}{3}\pi\right) 
 \Gamma\left( \frac{m+1}{3}\right)
 { \left( \frac{X+x}{6}\right)^{-\frac{m+1}{3}}} 
\label{AsymptExp}
\end{equation}
extending the expansion of Eq.~\eqref{Expans1}. (A proof of \eqref{AsymptExp} can be found in the treatise \cite{watson} which also contains explicit expressions for $B_m(x)$, polynomials in $x$ of order $m$, for small orders \cite{Notewatson}.) 
{}From Eq.~(\ref{AsymptExp}) we deduce that 
\begin{equation}
\label{V-smallbis}
v  =    {\mathcal A}  \,\lambda^{1/3} - \frac{\lambda}{10} + \frac{{\mathcal A}^2}{140} \,\lambda^{5/3} +
 {\mathcal O}\left( \lambda^2 \right) \, , 
\end{equation}
with ${\mathcal A}$ defined in Eq.~\eqref{V-final}. This expression is more precise than \eqref{V-final} and it can be used for data fitting (see figure~\ref{figuremath1}).

\subsection{Diffusion constant and higher cumulants}

The calculation of the velocity described in subsection~\ref{vel} relies on the steady-state probability distributions. The diffusion constant, however, cannot be determined from the steady-state distributions. Indeed, dispersion embodies information about transient states --- it can be expressed in terms of integrals of two-time correlation functions \cite{DEM}, while the knowledge of the steady-state characteristics is insufficient. Therefore in many interacting stochastic processes the analytical determination of the diffusion constant is quite involved \cite{DEM, DM, spiders}. In this subsection we describe an approach that leads to an analytical expression for the diffusion constant  ${\mathcal D}$. This approach will also allow us to compute the large deviation function which, in addition to velocity and diffusion constant, contains the entire statistics of the translocated length.  

Before embarking into calculations we notice that in Ref.~\cite{chou}, the diffusion constant has been probed through Monte-Carlo simulations. The value  ${\mathcal D} \approx 0.388$ was found numerically in the limit of vanishing deposition rate  $\lambda \ll 1$, whereas in the $\lambda \to \infty$ limit the model becomes equivalent to the one-dimensional burnt-bridge model \cite{may}, for which the dispersion has been calculated in \cite{bridge}. We now show how to compute $\mathcal{D}(\lambda)$ analytically. 

As explained in Section~\ref{sec:general}, we are interested in the statistical
properties of the translocated length  $L$  in the long time limit, see \eqref{L-av-var}. Since the motion of the polymer is affected only by the leftmost chaperone molecule, let us disregard the positions of other chaperon molecules and focus on $P_j(L,t)$, the probability that at time $t$ the polymer has translocated a 
length  $L$ inside the cell and that  the leftmost chaperone is at distance $j$ from the pore. The evolution equations for  $P_j(L,t)$ read
\begin{eqnarray}
 \frac{d P_j(L)}{dt} &=&  P_{j-1}(L-1) +  P_{j+1}(L+1) + 
 \lambda \sum_{k>j} P_k(L)
 - [2+\lambda(j-1)] P_j(L)   \,\,\, \hbox{ for } \, j>1,  \,
 \label{eq:PLj} \\
  \frac{d P_1(L)}{dt} &=&   P_2(L+1) +  \lambda \sum_{k>1} P_k(L)
   - P_1(L) \, .
  \label{eq:PL1}
\end{eqnarray}
These equations are just a more comprehensive version of  
Eqs.~\eqref{eq:Qj}--\eqref{eq:Q1}. Using the sum rule
$\sum_{L\geq j} P_j(L,t) = Q_j(t)$ which follows from the definitions of 
$P_j(L,t)$ and $Q_j(t)$ one can recover
Eqs.~\eqref{eq:Qj}--\eqref{eq:Q1} from Eqs.~\eqref{eq:PLj}--\eqref{eq:PL1}. 

We now introduce the generating 
function (which is essentially the discrete Laplace transform) 
\begin{equation}
     \Pi_j(\mu, t) = \sum_L { e}^{\mu L} P_j(L,t) 
 \label{def:Pi}
\end{equation}
In equations \eqref{eq:PLj}--\eqref{eq:PL1} and hereinafter we usually suppress the dependence on time, e.g. we shortly write $P_j(L)$ instead of $P_j(L,t)$. 
In addition to suppressing time dependence, we shall often suppress the dependence on the fugacity parameter $\mu$; for instance, $\Pi_j$ means 
$\Pi_j(\mu, t)$.
  
{}From Eqs.~\eqref{eq:PLj}--\eqref{eq:PL1} we deduce the 
governing equations for $\Pi_j$'s:
\begin{eqnarray}
 \frac{d \Pi_j}{dt} &=& 
 { e}^{\mu} \Pi_{j-1} + { e}^{-\mu} \Pi_{j+1}
  + \lambda \sum_{k>j} \Pi_k
 -\left[2+\lambda(j-1) \right]  \Pi_j
 \, \hbox{ for } \,\,\, j>1, 
 \label{eq:Pij} \\
  \frac{d \Pi_1}{dt} &=&  { e}^{-\mu} \Pi_2+ 
  \lambda \sum_{k>1}  \Pi_k  - \Pi_1 \, .
  \label{eq:Pi1}
\end{eqnarray}
Note that these equations are the same as  Eqs.~(\ref{eq:Qj}) and (\ref{eq:Q1})  
satisfied by  the $Q_j$'s,  except for the exp$(\pm\mu)$ factors. The infinite system of equations \eqref{eq:Pij}--\eqref{eq:Pi1} can be rewritten  in a matrix form
\begin{equation}
 \frac{d \boldsymbol{\Pi}(\mu)}{dt} =\mathbb{M}(\mu) \boldsymbol{\Pi}(\mu)\,.
\end{equation}  
Here $\boldsymbol{\Pi}(\mu)$ is a column vector
$\boldsymbol{\Pi}(\mu)=(\Pi_1,\Pi_2,\ldots)^{T}$ and $\mathbb{M}(\mu)$ is a matrix
\begin{equation}
  \mathbb{M}(\mu)  =  \begin{pmatrix} -1 & \,\,\,  { e}^{-\mu} + \lambda &
  \,\,\,  \lambda &   \,\,\,  \lambda  & \lambda & \lambda & \dots &\\
\,\,\,   { e}^{\mu} & -(2+\lambda) & \,\,\,{ e}^{-\mu} + \lambda &
   \,\,\,  \lambda &  \lambda  & \lambda & \dots \\
 \,\,\, 0& \,\,\,    { e}^{\mu}  & -(2 +2\lambda ) &
 \,\,\, { e}^{-\mu} + \lambda & \lambda   &  \lambda  & \dots  \\
&\,\,\,0& \,\,\,{ e}^{\mu} &-(2+3\lambda )
  & \,  { e}^{-\mu} + \lambda&  \,\,\,  \lambda& \dots  \\
& &  \ddots & \ddots &  & \ddots  \,\,\,&  \,\,\,\ddots  \\
  \end{pmatrix}
\label{eq:defMmu}
\end{equation}  
For $\mu =0$,  the matrix  $\mathbb{M}(0)$ is the  continuous-time Markov matrix
that governs the evolution  of the  $Q_j$'s. For non-vanishing  $\mu$,
we can write 
\begin{equation}
 \langle  { e}^{\mu L}  \rangle = \sum_j \Pi_j(\mu,t) \sim
  { e}^{U(\mu) \,  t } \quad\text{for}\quad t\to\infty
\label{def:Umax}
\end{equation} 
with $U(\mu)$ being the largest  eigenvalue of matrix $\mathbb{M}(\mu)$.
(The existence and the uniqueness  of the maximal eigenvalue
is guaranteed by the Perron-Frobenius theorem \cite{vankampen}.)   
In the long time-limit, the function $U(\mu)$  generates all  the cumulants of 
$L$. Indeed, taking the logarithm of Eq.~(\ref{def:Umax}), and expanding it
with respect to $\mu$, one obtains
\begin{equation}
 U(\mu) = \mu \frac{ \langle L  \rangle }{t}
 + \mu^2 \frac{ \langle L^2 \rangle - \langle L  \rangle^2}{2t} +\ldots = 
 \mu v +  \mu^2 {\mathcal D} + \ldots
\label{dvpteigenval}
\end{equation} 
Hence, $U(\mu)$ contains the full information for the long-time statistics
of the translocation length; in particular, the second-order term in the Taylor expansion provides  the diffusion constant.

An implicit  formula  for $U(\mu)$ can be derived following a path 
very similar to that  used in the previous section to calculate the velocity $v$.
The computations are more cumbersome [see Appendix~\ref{App:1}] but
straightforward. We obtain 
 \begin{equation}
U(\mu) =  ({ e}^{\mu} - 1)
 \left( 1 -   \frac{ J_{\Lambda[1 + U(\mu)/2]   +1}(\Lambda)}
 {J_{ \Lambda[1 +U(\mu)/2]}(\Lambda)}  \right)  
\,\,\, \hbox{ with } \,\,\,  \Lambda =\frac{2}{\lambda} \,.
\label{sol:Umax}
\end{equation}
A perturbative expansion for small $\mu$  allows  us  to calculate
the cumulants of the translocated length $L$. At the first order we have:
 \begin{equation}
 \label{U_vD}
U(\mu) =  \mu  \left( 1 -   \frac{ J_{\Lambda   +1}(\Lambda)}
 {J_{ \Lambda}(\Lambda)}  \right) + {\mathcal O}\!\big(\mu^2 \big),
\end{equation}
which agrees with the already-known expression \eqref{eq:vitesse}
for the translocation velocity.
Developing to the second order allows us to derive a formula for the
diffusion constant of the polymer chain
\begin{equation}
  {\mathcal D} = \frac{v}{2} - \frac{v}{\lambda}
  \Big( \frac{\partial}{\partial \Lambda}
  \frac{ J_{\Lambda+1}(x)} {J_{ \Lambda}(x)} \Big) \Big|_{x =\Lambda} \, . 
 \label{formule:D}
\end{equation}
We emphasize  that the derivative is  with respect to the order $\Lambda$
of the Bessel function and not  with respect to  its  argument $x$, 
i.e. the value $x =\Lambda$ is to be substituted after the derivative has been taken. 
 Using  the asymptotic expansion~(\ref{AsymptExp}), we deduce 
\begin{equation}
\label{D-small}
{\mathcal D}  =   {\mathcal A}^3  +   \frac{{\mathcal A}^2}{30} \,\lambda^{2/3} +
 {\mathcal O}\!\big( \lambda^{4/3} \big) 
\end{equation}
with ${\mathcal A}$ defined by Eq.~\eqref{V-final}.
The exact formula~\eqref{formule:D} and the small
 $\lambda$ expansion are represented in 
 figure~\ref{figuremath1}.
In the leading order,  we find 
\begin{equation}
{\mathcal D} \simeq   {\mathcal A}^3 =  3
 \left(  \frac{\Gamma(2/3)}{\Gamma(1/3)}\right)^3 = 0.3874382381\ldots 
 \label{eq:Ddiscrete}
\end{equation}
The Monte-Carlo simulation results \cite{chou} are in excellent agreement 
with this analytical prediction. At first sight, the agreement even seems too good: Equation \eqref{D-small} shows that the corrections to the limiting value are of the order $\lambda^{2/3}$, so they could be significant  even for small values of the attachment rate $\lambda$. However, the amplitude in front of the correction
term is very small (numerically 0.00851). This explains why  in numerical
simulations \cite{chou}  the dispersion looks  almost constant for $\lambda < 1$.

Thus we have obtained the complete statistics of the translocated length of the polymer in the situation when the chaperone molecules have the same length as the monomer units. Our chief result, Eq.~\eqref{sol:Umax}, is an implicit highly transcendental relation for $U(\mu)$, the largest  eigenvalue of matrix 
$\mathbb{M}(\mu)$. The generating 
 function $U(\mu)$ is related by Legendre Transform to the large deviation function 
 of the translocated length $L$
 (see e.g. Ref.~\cite{LD} on large deviation techniques). We emphasize again that
  this generating  function contains {\em all} cumulants: expanding this  function, 
$U(\mu)=\sum_{n\geq 1}\mu^n \tfrac{U_n}{n!}$, we extract $v=U_1$,
$\mathcal{D}=2U_2$, see \eqref{U_vD}, and subsequent cumulants 
\begin{equation*}
\begin{split}
U_3 &= \lim_{t\to\infty}\frac{\langle L^3\rangle
             -3\langle L\rangle \langle L^2\rangle + 2\langle L\rangle^3}{t}\,,\\
U_4 &= \lim_{t\to\infty}
\frac{\langle L^4\rangle
             -4\langle L\rangle \langle L^4\rangle  - 3\langle L^2\rangle^2
             +6 \langle L\rangle^2 \langle L^2\rangle}{t}\,, \,\,\text{etc.}
\end{split}
\end{equation*}

Finally we note that all results of this section can be generalized to the situation when the bare polymer undergoes a biased diffusion; an outline of the computations and the final formulas for the velocity and the diffusion constant are presented in Appendix~\ref{App:asym}.  
        
\section{The continuous space model}
 \label{sec:continuous}
 
A class of proteins responsible for the translocation mechanism  across the
endoplasmic reticulum is known as the Hsp-70 family. Typical chaperones
belonging to that family are of 2nm in size,  representing  roughly 6 or 7
lattice sites of the  translocating polymer. The size $\ell$  of the  chaperones 
can have  an important quantitative  effect on the translocation process,
because   it modifies  the velocity $v$ and the diffusion constant ${\mathcal D}$,
as explained in Eq.~(\ref{dim}).  For chaperones of length $\ell$  greater than  
the  monomer  size $a$, the discrete model of the previous section can in principle be solved. The resulting expressions for the velocity are not explicit, however, 
because a linear system  of $\ell$ equations must be solved in order to adjust
the boundary conditions \cite{chou}. Here we examine the limiting case when the length of the chaperone molecule vastly exceeds the monomer size. Keeping the 
chaperone  of size $\ell$ finite we therefore consider the $a\to 0$ limit when the discrete lattice turns into a continuous line. 

We set again the length $\ell$  of the  chaperone protein  and the bare diffusion constant $D$  to unity. With this agreement, the dimensionless attachment rate $\lambda \ell^3/D$ is equal to $\lambda$. Thus mathematically the chaperon molecules are segments of unit length that absorb onto a continuous line with rate $\lambda$. Adsorption problems where the substrate is the continuous line are usually solvable if the analogous problems where the substrate is  
the one-dimensional lattice are tractable \cite{PaulK}. 
 
\begin{figure}[ht]
\begin{center}
\includegraphics[width=0.3\textwidth,angle=-90]{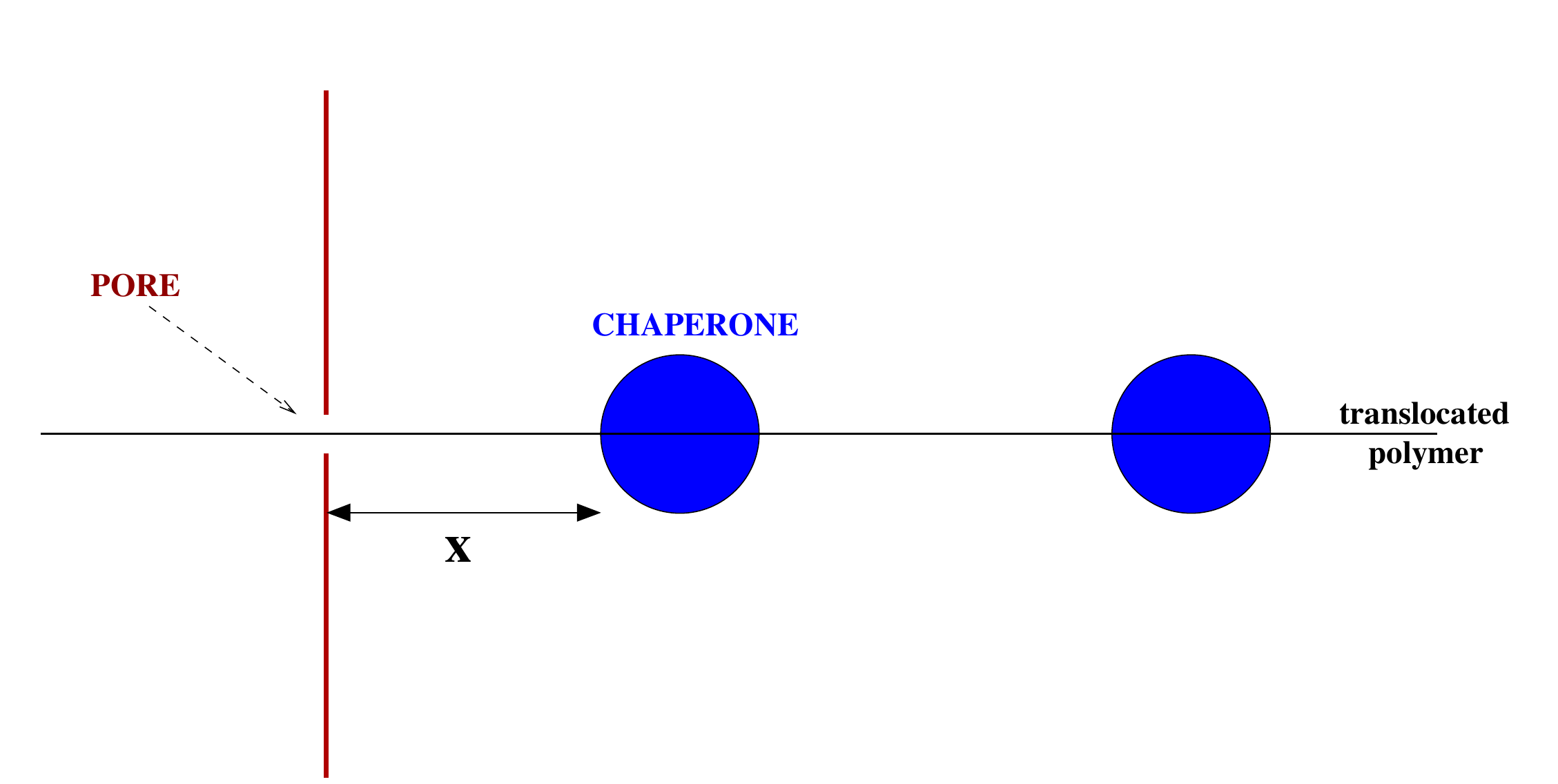}
\caption{Illustration of chaperon-assisted translocation for the continuous model.}
\label{fig-transloCont}
\end{center}
\end{figure}

Let $Q(x,t) dx$ be the probability that  the left edge of the  first 
absorbed chaperone on the right of the pore  is located between $x$ and $x+dx$,
see figure~\ref{fig-transloCont}.  Further, denote by $E(x,t)$ 
the probability that the first chaperone molecule is 
more than a  distance $x$ away. The relation between these two probabilities is
\begin{equation}
E(x,t) = \int_x^\infty dy\,Q(y,t)
\end{equation}

The probability densities $Q(x,t)$  satisfy
\begin{subequations}
\begin{align}
&\frac{\partial Q}{\partial t} = \frac{\partial^2 Q}{\partial x^2} -\lambda(x-1)Q
+\lambda\int_{x+1}^\infty dy\,Q(y),  \quad x \ge 1
\label{large}\\
&\frac{\partial Q}{\partial t} =\frac{\partial^2 Q}{\partial x^2} 
+\lambda\int_{x+1}^\infty dy\,Q(y), \quad x<1
\label{small}
\end{align}
\end{subequations}
These equations can be derived directly by studying how  $Q(x,t)$
 varies between time $t$ and $t+dt$. One can also obtain them as 
 continuous limits of the discrete equations~\eqref{eq:Qj}
 and \eqref{eq:Q1}.

The governing equations for the empty interval probabilities $E(x,t)$ are
\begin{subequations}
\begin{align}
&\frac{\partial E}{\partial t} = \frac{\partial^2 E}{\partial x^2} -\lambda(x-1)E
-\lambda\int_x^{x+1} dy\,E(y),  \quad x \ge 1
\label{large_E}\\
&\frac{\partial E}{\partial t} =\frac{\partial^2 E}{\partial x^2} 
-\lambda\int_1^{x+1} dy\,E(y), \quad x<1
\label{small_E}
\end{align}
\end{subequations}
In the large time limit,  $Q(x,t)$ and  $E(x,t)$ become stationary. Here we focus on the stationary empty interval probabilities  $E(x)\equiv E(x,t=\infty)$ that satisfy
\begin{subequations}
\begin{align}
&\frac{d^2 E}{d x^2} = \lambda(x-1)E(x)
+\lambda\int_x^{x+1} dy\,E(y),  \quad x \ge 1
\label{large_stat}\\
&\frac{d^2 E}{d x^2} =
\lambda\int_1^{x+1} dy\,E(y), \quad x<1
\label{small_stat}
\end{align}
\end{subequations}
This set of integro-differential equations can be viewed as a generalization
of the classical Airy equation \cite{Airy}. Essentially, we need to solve \eqref{large_stat};  the solution of \eqref{small_stat} will  then be found by a simple integration
as the right-hand side is known. Equation \eqref{large_stat} 
is non-local  as can be seen  by  differentiating it with respect to $x$:
\begin{equation*}
\frac{d^3 E(x)}{d x^3} = \lambda (x-2)\,\frac{d E(x)}{d x} + \lambda E(x+1)
\end{equation*}
In the following, we derive an exact solution of 
Eqs.~\eqref{large_stat}--\eqref{small_stat}.

\subsection{Stationary solution and velocity}

For difference-differential equations, it is difficult even
to prescribe the boundary conditions; formally one needs
to have a boundary condition on the interval of length 1,
say $E(x)$ on interval $1\leq x\leq 2$, and then one can 
construct the solution for all $x>1$. For almost any choice
of the `boundary values' the resulting solution will be unacceptable
(diverging at $x\to\infty$ rather than vanishing, not decreasing monotonously  with $x$, etc...). Rather than proceeding in a formal manner, we shall use  here 
the Laplace method (see e.g. \cite{landau} or \cite{goursat}) which is a very powerful approach allowing one to deal with linear differential equations whenever coefficients are linear functions of $x$. This constraint is met in our case, so we employ the Laplace method.  This method tells one to seek a solution in the form
\begin{equation}
\label{ansatz}
E(x) = \int_C du\,e^{ux} Z(u)
\end{equation}
where the integral is taken over a contour $C$ in a complex plane to be specified
at a later stage. 
 
\begin{figure}[ht]
\begin{center}
\includegraphics[width=0.35\textwidth,angle=-90]{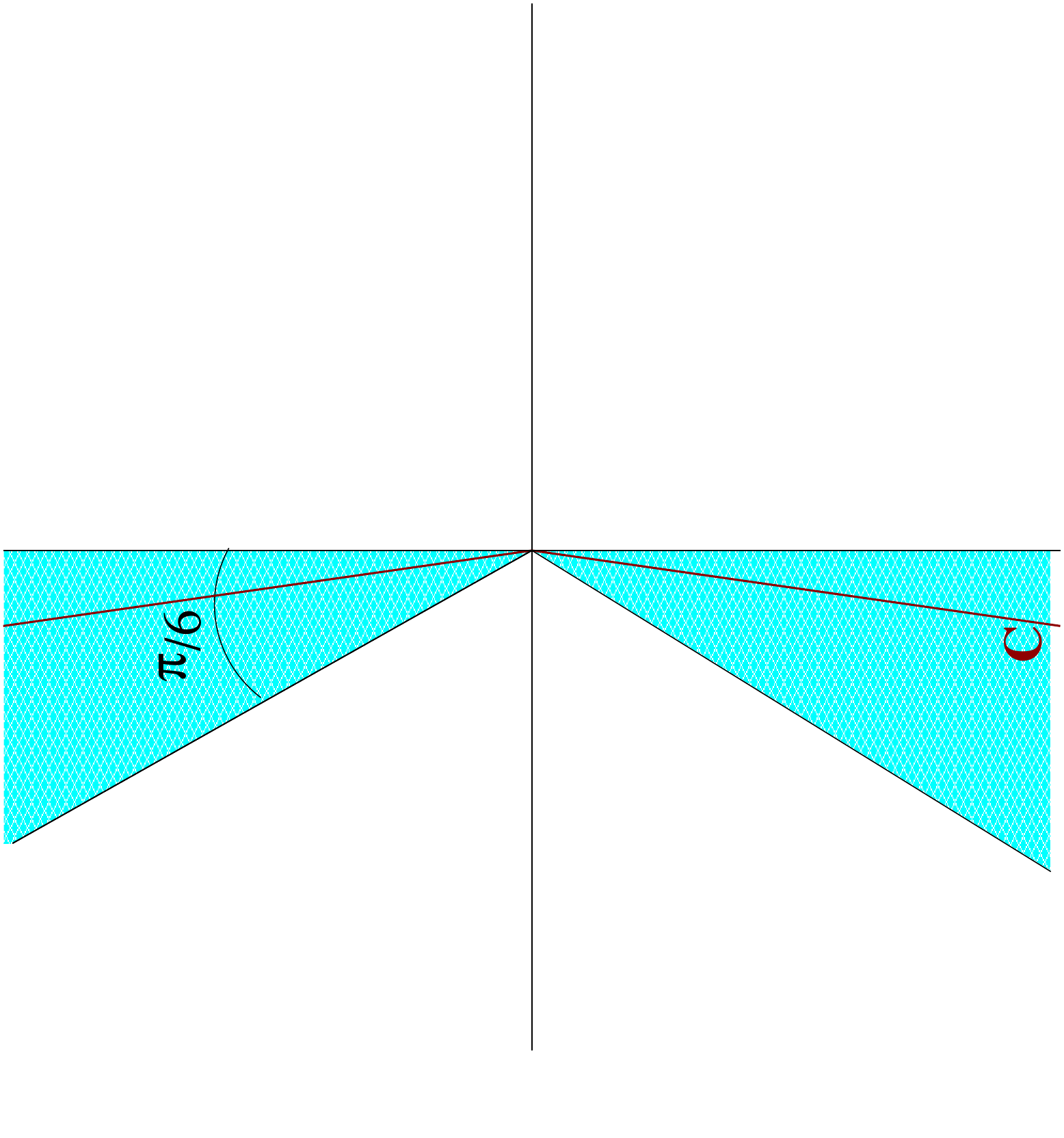}
\caption{Contour  in the complex plane which has been used in the 
implementation of Laplace's method in Eq.~\eqref{ansatz}. The shaded domain
 (with opening angle equal to $\pi/6$)  shows the region 
  in which the contour $C$ can exist
 and leads to a well defined function $Z(u).$ }
\label{Fig:CLaplace}
\end{center}
\end{figure}

We determine  $Z(u)$  by  inserting  \eqref{ansatz} into the governing 
equation  \eqref{large_stat}. The left-hand side is
\begin{equation*}
\frac{d^2 E}{d x^2} = \int_C du\,e^{ux} u^2 Z(u)
\end{equation*}
The non-local term on the right-hand side becomes
\begin{equation*}
\int_x^{x+1} dy\,E(y) = \int_C du\,e^{ux}\, \frac{e^u-1}{u}\, Z(u)
\end{equation*}
To simplify $xE(x)$ we use integration by parts 
\begin{eqnarray*}
xE(x) &=&\int_C du\,e^{ux} x Z(u)\\
&=& e^{ux} Z(u)\big|_C - \int_C du\,e^{ux} \frac{dZ}{du}\\
&=& - \int_C du\,e^{ux} \frac{dZ}{du}
\end{eqnarray*}
where the validity of the last result relies on the assumption that function
$e^{ux} Z(u)$ has the same value on both ends of the contour $C$. 
Combining above results we find that \eqref{ansatz} is 
a solution to Eq.~\eqref{large_stat} if $Z(u)$ obeys 
\begin{equation*}
u^2 Z = -\lambda Z - \lambda\, \frac{dZ}{du} + \lambda\, \frac{e^u-1}{u}\, Z
\end{equation*}
Solving this differential equation we find
\begin{equation}
\label{Zu-sol}
Z = {\rm const}\cdot
\exp\!\left(-u - \frac{u^3}{3\lambda} + \int_0^u dv\,\frac{e^v-1}{v}\right)
\end{equation}
If we take the contour $C$ that goes from infinity to the
 origin along the ray with ${\rm arg}(u)=\frac{3\pi}{2}-\epsilon$ 
and then goes back to infinity along  the ray 
with ${\rm arg}(u)=\frac{\pi}{2}+\epsilon$ (see figure~\ref{Fig:CLaplace}), then 
 $e^{ux} Z(u)$ vanishes on both ends of the contour
 $C$ (as long as $0<\epsilon<\pi/6$). In the $\epsilon\to +0$ limit,
 the contour $C$ coincides with  the imaginary axis. Writing $u=iw$ we 
 obtain  the desired integral representation 
\begin{equation}
\label{E-sol}
E(x) = A\int_0^\infty dw\,C(w)\,C(w,x)
\end{equation}
with 
\begin{eqnarray}
\label{sol-C}
C(w)       &=&  \exp\!\left(\int_0^w dv\,\frac{\cos v-1}{v}\right)\nonumber\\
C(w,x)    &=& \cos[w(x-1)+W]\\
W=W(w) &=& \frac{w^3}{3\lambda} + \int_0^w dv\,\frac{\sin v}{v}\nonumber
\end{eqnarray}
and yet undetermined constant $A$. Equation \eqref{E-sol} is valid for $x \ge 1$ and can be used to determine the behavior of $E(x)$ at large $x$ (see Appendix~\ref{app:largex}).

The above solution \eqref{ansatz}--\eqref{Zu-sol} is valid  for $x>1$. Using these results we reduce equation \eqref{small_stat} governing the small size ($x<1$) behavior to 
\begin{equation}
\label{E-small}
\frac{d^2 E}{d x^2} = \lambda A\int_0^\infty dw\,C(w)\,
\frac{\sin(wx+W)-\sin(W)}{w}\,.
\end{equation}
Integrating \eqref{E-small} subject to $E(0)=1$ we obtain
\begin{eqnarray}
\label{E-small-sol}
E &=&  1-Bx- \lambda A\int_0^\infty dw\,C(w)\Phi(w,x) \\
\hbox{ with } \quad 
 \Phi &=& \frac{\sin(wx+W)-\sin W -xw\cos W }{w^3}+x^2\,\frac{\sin W}{2w} \, .
 \label{def:phi}
\end{eqnarray}
The constants $A$ and $B$ are determined  using the fact that 
the probability $E(x)$ is continuous, and therefore the two expressions, 
\eqref{E-sol} and \eqref{E-small-sol}, must match at  $x=1$.  This gives
\begin{eqnarray}
\label{AB1}
\frac{1-B}{A}&=&\int_0^\infty dw\,\mathcal{W}_1(w)\,C(w) \\
\hbox{ with } \quad \quad
\mathcal{W}_1 &=&\cos W +\lambda\,
\frac{\sin(w+W)-\big(1-\tfrac{w^2}{2}\big)\sin W-w\cos W}{w^3}
 \label{def:W1}
\end{eqnarray}
We should further equate the derivatives of $E(x)$ at the
 matching point. This yields 
\begin{eqnarray}
\label{AB2}
\frac{B}{A}&=&\int_0^\infty dw\,\mathcal{W}_2(w)\,C(w) \\
\hbox{ with } \quad  
\mathcal{W}_2 &=&\  w\,\sin W -\lambda\,\frac{\cos(w+W)
 -\cos W +w\sin W}{w^2} \, .
 \label{def:W2}
\end{eqnarray}

\begin{figure}[ht]
\begin{center}
\includegraphics[width=0.5\textwidth]{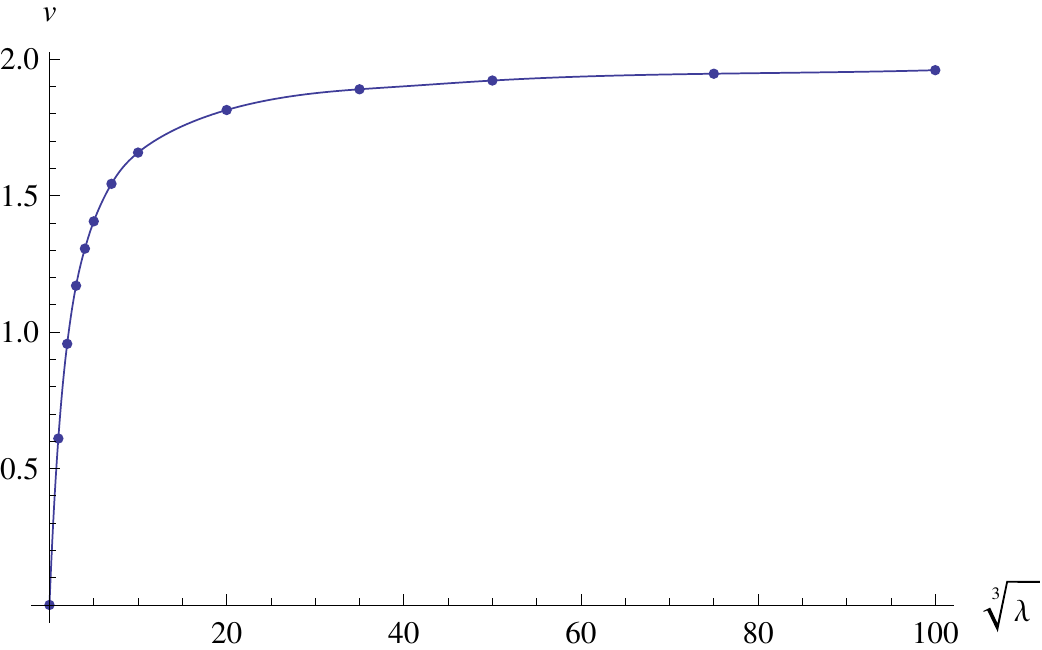}
\caption{Plot of the velocity $v$  of the continuous model
  as a function of $\lambda^{1/3}$
  using the exact formula~\eqref{eq:Formulev}. The integrals that appear in
 the numerator and denominator of  expression~\eqref{eq:Formulev}
 were evaluated for various values of $\lambda$ ranging between 1 and
 $10^6$ (represented as dots in the figure) and a  smooth interpolating
 curve was drawn to join these points. We observe that for large
 values of $\lambda$, we have $v \to 2$. An empirical approximation,
 valid for $\lambda^{1/3} \ge 10$,  is given by $v = 2
 - \frac{c_1}{\lambda^{1/3}}
 + \frac{c_2} {\lambda^{2/3}}...$ with $c_1 \simeq 4$ and $c_2 \simeq 6.$ }
\label{figuremath2}
\end{center}
\end{figure}

Equation~\eqref{E-sol} and Eqs.~\eqref{E-small-sol}--\eqref{def:W2} 
determine the  probabilities  $E(x)$   in the stationary regime. The knowledge of these stationary empty interval probabilities allows us to compute the velocity. Using 
$v= -E'(0)$, we deduce from \eqref{E-small-sol}  that $v=B$. [The relation $v= -E'(0)$ is the continuous version of \eqref{eq:vitesse}.] Equations \eqref{AB1}--\eqref{AB2} in conjunction with $v=B$ lead to the following expression for the velocity:
\begin{equation}
v= \frac{\int_0^\infty dw\,\mathcal{W}_2(w)\,C(w)}
{\int_0^\infty dw\,[\mathcal{W}_1(w)+\mathcal{W}_2(w)]\,C(w)} \, .
\label{eq:Formulev}
\end{equation}
This formula can be used to plot  $v$ as a function of the attachment
rate $\lambda$, see figure~\ref{figuremath2}, 
 and to extract asymptotic expansions. In the case 
$\lambda \to 0,$ we make the change of variable $ w = \lambda^{1/3} \omega$ 
in Eq.~\eqref{eq:Formulev}. Taking the $\lambda \to 0$ limit and keeping $\omega$ finite,  we find that the dominant  behaviors of the
expressions in Eqs.~\eqref{sol-C}, \eqref{def:W1} and~\eqref{def:W2} are given by 
\begin{equation}
 C=1, \quad   W=\tfrac{ \omega^3}{3}, \quad
 \mathcal{W}_1 = \cos\!\big(  \tfrac{ \omega^3}{3} \big) \quad \hbox{and}
\quad  \mathcal{W}_2=\lambda^{1/3} \omega \sin\!\big(\tfrac{ \omega^3}{3} \big)\, .
\label{eq:dominantlambda0}
\end{equation}
Substituting in Eq.~\eqref{eq:Formulev}, we obtain the leading behavior of the speed 
\begin{equation}
  v = \lambda^{1/3} 
\frac{\int_0^\infty d\omega\, \omega \sin\!\big(\tfrac{ \omega^3}{3} \big)}
    {\int_0^\infty d\omega\,\cos\!\big(  \tfrac{ \omega^3}{3} \big)} = 
 \frac{3^{1/3} \Gamma(2/3)}{\Gamma(1/3)} \lambda^{1/3}  \, . 
 \label{eq:vsmalllambda}
\end{equation}
This result agrees with the prediction for the discrete model, 
Eq.~\eqref{V-final}. This agreement is not surprising --- in the limit of infinitely small attachment rate, the discreteness of the polymer does not play any role and the system behaves as if it were effectively continuous.
 
The  $\lambda \to \infty$  limit is more difficult  to implement (see Appendix~\ref{App:deformed-operator}), yet the final result is remarkably simple
\begin{equation}
  v \to 2   \quad \hbox{when}  \quad \lambda \to \infty \, . 
  \label{limitspeedLlambda}
\end{equation}
We can understand \eqref{limitspeedLlambda} as follows. In the 
$\lambda \to \infty$ limit,  a chaperone is attached   as soon as the
translocated segment  length  reaches 1. The  model becomes equivalent to a  continuous time random walker  on a discrete lattice, that can only make forward steps of unit length. The randomness in the problem arises from the distribution 
$\psi(T)$ of the  waiting time $T$  between two successive jumps. The distribution  $\psi(T)$ is identical to that
of the first passage time at $x=1$  of a Brownian motion starting at $x=0$ with a reflecting 
 boundary at  $x=0$ and an absorbing  boundary at  $x=1$. The Laplace transform 
 $\hat\psi(p) = \langle  { e}^{pT} \rangle$ is obtained by utilizing 
 standard techniques \cite{Montroll,Sid,PaulK} to yield
\begin{equation}
  \hat\psi(p) =  \frac{1}{\cosh\sqrt{p}} \,.
 \label{distrbWT}
\end{equation}
The velocity is then found to be
 $$ v = \frac{1}{ \langle T \rangle} = -\frac{1}{\hat\psi'(0)} = 2\,.$$

 \subsection{Calculation of the diffusion constant}

The diffusion constant ${\mathcal D}$ of the continuous translocation model can be derived by utilizing the same approach as in the discrete case. Namely, we shall establish an implicit equation for the large deviation function $U(\mu)$. Mathematically, one has to find the largest eigenvalue of a suitable  differential operator  ${\mathcal L}_\mu$ which is a deformation (with respect to the  parameter $\mu$) of the evolution operator for $E$ that appears in Eqs.~\eqref{eq:evolEgoth}. One also has to be careful of how the boundary conditions are deformed. 

The details of the calculations are given in Appendix~\ref{App:deformed-operator}  and the solution for $U(\mu)$  in presented in Eq.~\eqref{eq:SolUmuimplicit}.  Having computed $U(\mu)$,  we can retrieve the formula~\eqref{eq:Formulev} for the velocity (by expanding   $U(\mu)$  to the first order with respect to $\mu$). The coefficient of the $\mu^2$ term gives the  diffusion constant ${\mathcal D}$. The calculations are rather involved but systematic. The explicit expression for 
${\mathcal D}$ is given in Eqs.~\eqref{eq:formuleDcont}
 to \eqref{formuledeltaW2}. Although the results look unwieldy, they can be used to derive asymptotic limits  for very small or very large values of the attachment rate $\lambda$.

In the case  $\lambda \to 0$ we make again the change of variable 
$w = \lambda^{1/3} \omega$. Using 
\eqref{eq:dominantlambda0}--\eqref{eq:vsmalllambda} we find the leading behavior of the functions $\delta{C}, \delta\mathcal{W}_1, \delta\mathcal{W}_2$ (see Appendix~\ref{App:deformed-operator} for definitions) 
 \begin{eqnarray}
 \label{delta_CWW}
   \delta{C} = - \frac{{\omega}^2}{\lambda^{1/3}} \,, \quad
    \delta\mathcal{W}_1 = - \frac{ {\mathcal A}} {\lambda^{1/3}}\, {\omega}\,
     \sin\!\big(\tfrac{ \omega^3}{3} \big) 
   \,, \quad \delta\mathcal{W}_2 = {\mathcal A} {\omega}^2 
   \cos\!\big(  \tfrac{ \omega^3}{3} \big) \, .
\end{eqnarray}
{}From this information we extract the limiting behavior of ${\mathcal D}$  by analyzing each contribution in Eq.~\eqref{eq:formuleDcont}: 
\begin{equation}
{\mathcal D} \simeq  1 - v  \frac{\int_0^\infty dw\,C  \delta\mathcal{W}_1   }
{\int_0^\infty dw\,[\mathcal{W}_1 +\mathcal{W}_2]\,C} +
 \frac{\int_0^\infty dw\,\mathcal{W}_2 \,\delta C  }
{\int_0^\infty  dw\,[\mathcal{W}_1 +\mathcal{W}_2 ]\,C}  +o(1)
\end{equation}
Substituting the  dominant behaviors found above,
\eqref{eq:dominantlambda0}--\eqref{eq:vsmalllambda} and \eqref{delta_CWW},
we arrive at
\begin{equation}
{\mathcal D} \simeq 1 +   {\mathcal A}^2\,
  \frac{\int_0^\infty d{\omega}\, {\omega}\,\sin\! \frac{ \omega^3}{3} }
  {\int_0^\infty d{\omega}\, \cos\! \frac{ \omega^3}{3} } - 
  \frac{\int_0^\infty d{\omega}\, {\omega}^3\sin\! \frac{ \omega^3}{3} }
  {\int_0^\infty d{\omega}\, \cos\!\frac{ \omega^3}{3} } \, . 
\end{equation}
Calculating the integrals in the above formula (the numerator in the third
term is determined using analytic continuation and we find that  the  third term is equal to -1),
  we obtain, in agreement
with  the result of Eq.~\eqref{eq:Ddiscrete} for the discrete case,  
 \begin{equation}
   {\mathcal D} \simeq  {\mathcal A}^3   \quad \hbox{when}  \quad \lambda \to  0\, .
  \label{limitdiffusionsmalllambda}
\end{equation}

The case  $\lambda \to \infty$ is analyzed using  techniques and tricks similar to those  involved  for deriving the large  $\lambda$ asymptotics in the discrete case
(see Appendix~\ref{App:deformed-operator} for details).  A precise investigation of the  asymptotic behavior of each of the terms that appear in the formula~\eqref{eq:formuleDcont} for the diffusion constant can be carried out. The calculations are systematic and in principle straightforward, but very lengthy;
the final result is shockingly compact
  \begin{equation}
   {\mathcal D} \to \frac{2}{3}   \quad \hbox{when}  \quad \lambda \to \infty \, . 
 \label{limitdiffusionLlambda}
\end{equation}
This value is in perfect agreement with the one obtained for the  continuous time random walker model with waiting time  $T$ distributed according to the law  $\psi(T)$  with Laplace transform $\hat\psi(p)$ given in Eq.~\eqref{distrbWT}. For this effective problem, the diffusion constant is found to be
 $$ {\mathcal D}  = \frac{1}{2} \frac{\langle T^2 \rangle - \langle T \rangle^2  }
{ \langle T \rangle^3} = \frac{1}{2} 
 \frac{\hat\psi''(0) - [\hat\psi'(0)]^2}{[-\hat\psi'(0)]^3} = \frac{2}{3}\,. $$
We remark that for very large values of the attachement rate $\lambda$,   the chaperon-assisted translocation model becomes equivalent to the burnt bridge model with periodic bridges \cite{bridge}. The velocity and diffusion coefficient for this model are known and using these results it is possible to determine the velocity and the diffusion coefficient for the chaperon-assisted translocation in the $\lambda  \to \infty$ limit. The connection with the burnt bridge model with periodic bridges holds even in the general discrete case with arbitrary $\ell/a$ \cite{chou}; if $\ell/a \to \infty$, the limiting values $v=2$ and ${\mathcal D}  = 2/3$ are recovered. Finally we note that the generating function technique employed in this study can be adapted to a calculation of the large deviation function of the burnt bridge model.

\section{Discussion}
 \label{sec:discussion}
 
The mechanism that drives the translocation of a protein through a membrane depends on chemical asymmetries between the cis and trans sides  of the membrane \cite{PeskinPNAS}. There can be several mechanisms leading
to asymmetry and they can be simultaneously present.  In this work,
we have studied a model where asymmetry arises due to the preferential binding of
proteins on the translocated segment of the polymer. The mathematical effect that gives raise to  directed motion is the Brownian ratchet model that plays a crucial role in the field of molecular motors \cite{prost,reimann,lubensky,lau}. 
It is important to note that the ratchet paradigm  is quite insensitive to the precise
mechanism at the molecular level. The key  ingredients are the breaking of spatial symmetry and a supply of free energy which insures  here  a sufficiently
strong binding of the chaperone on the polymer. In biological context, the free energy source is usually provided by ATP hydrolysis \cite{PNelson,deDuve}. 

{}From the point of view of statistical physics, it is 
interesting to note that these models, which are effectively one-dimensional,
can be mapped to variants of driven diffusive gases and can be studied through
techniques which were developed to understand the physics of non-equilibrium systems. These methods  not only  reveal underlying time-reversal symmetries of the system, but also provide powerful computational tools.  For example, the technique based on the deformation of the master operator can be traced back to the  proof of the Gallavotti-Cohen fluctuation theorem for Langevin dynamics and for Markov processes by Kurchan \cite{kurchan} and by Lebowitz and Spohn \cite{spohn}. In the present work, we have assumed that the attachment of chaperones
is irreversible; if the detachment rate were positive, the fluctuation theorem would be valid.  The analysis carried out in this paper heavily relies on the fact
that the equations of motion depend only  on the position of the chaperone closest to the pore. If the chaperones could detach from the polymer, the above property would no longer be true and  the positions of all the chaperones bound at
a given time to the polymer must be taken into account. The system would turn into 
an $M-$body problem with both $M$ (the number of currently attached chaperon molecules) and $L$ (the current length of the translocated polymer) being random variables. This appears to be a very challenging unsolved model. 
Besides, similar descriptions appear in  mass transport models  within a fungal hypha
with mutually excluding particles progressing in a  stochastic manner 
along a growing one-dimensional lattice \cite{martin}.

\acknowledgments{We  are grateful to  Tibor Antal,  Sid Redner and  
 Shamlal Mallick for useful  discussions and   careful reading
  of the manuscript. PLK thanks NSF grant CCF-0829541 for support.}

\appendix

 \section{Calculation of the Cumulant Generating Function}
  \label{App:1}

In this Appendix, we show that  the maximal eigenvalue  $U(\mu)$ of   
matrix $\mathbb{M}(\mu)$, defined in Eq.~(\ref{eq:defMmu}), is indeed 
given by Eq.~\eqref{sol:Umax}. 

Let $G_1,G_2,\ldots$ be the components of the eigenvector corresponding to the maximal eigenvalue  $U(\mu)$.  The eigenvalue equation reads
\begin{eqnarray} 
 U(\mu) G_j &=&
  { e}^{\mu} G_{j-1} +  { e}^{-\mu} G_{j+1} + \lambda \sum_{k>j} G_k
 -\left[2+\lambda(j-1) \right] G_j 
 \, \hbox{ for } \,\,\, j>1, 
 \label{eq:Gj} \\
      U(\mu) G_1 &=&  { e}^{-\mu} G_2 + 
  \lambda \sum_{k>1} G_k - G_1  =  { e}^{\mu} G_0 + { e}^{-\mu} G_2 + 
  \lambda \sum_{k>1} G_k - 2G_1  \, , 
  \label{eq:G1}
\end{eqnarray}
where,  in order to fit  the   equation for $j=1$  into the general  pattern, we have defined $e^\mu G_0 = G_1$. As in the beginning of Section~\ref{sec:discrete}, it is useful to introduce  
 $\Gamma_m =  \sum_{k>m} G_k$, defined  for $m \ge -1 $.
 Then, for all $m \ge 0$,  the following equation is satisfied:
\begin{equation}
  U(\mu) \Gamma_m =  { e}^{-\mu}
\Gamma_{m+1} - \left(2+ \lambda m  \right)\Gamma_m + 
  { e}^{\mu}\Gamma_{m-1} 
   \, . \label{eq:Gamma}
\end{equation}
 The  value of  $\Gamma_0$  is fixed  by imposing the normalization 
constraint: $\Gamma_0 = 1$. (This overall proportionality constant  does  not affect  the final result.) Equation~(\ref{eq:Gamma}) for $m=0$ then gives
\begin{equation}
 { e}^{-\mu}\Gamma_{1} + 
  { e}^{\mu}\Gamma_{-1}  =  2 +  U(\mu) \, .
 \label{eq:BC1}
\end{equation}
In terms of the $\Gamma$'s,  the boundary condition $e^\mu G_0 = G_1$ becomes
\begin{equation}
  \Gamma_{-1} + \Gamma_{1}  { e}^{-\mu} = 1 +  { e}^{-\mu} \, .
  \label{eq:BC2}
\end{equation}
To solve  Eq.~(\ref{eq:Gamma})  with the boundary conditions~(\ref{eq:BC1}) and (\ref{eq:BC2}),  we  use the following ansatz: 
\begin{equation}
  \Gamma_{m} =  \frac{  J_{m + r} (x)}
 {J_{ r} (x)} { e}^{ m \mu}  \, ,  \label{eq:solGammam}
\end{equation}
where $r$ and $x$ have to be determined.
Combining Eqs.~(\ref{eq:Gamma}) and  \eqref{eq:solGammam}, we get 
\begin{equation}
   x  = 2/\lambda = \Lambda \,\, \hbox{ and } \,\,\
  r\lambda = 2 +  U(\mu) \, . 
\end{equation} 
 Hence
 \begin{equation}
 \Gamma_{m} =  \frac{  J_{m + r} (\Lambda)}
 {J_{ r} (\Lambda)} { e}^{ m \mu}  \,.
\label{sol:Gammam}
\end{equation} 
   The normalization $ \Gamma_{0} =1$ is satisfied. We eliminate
 $\Gamma_{-1}$ between Eqs.~(\ref{eq:BC1})
  and (\ref{eq:BC2}) and obtain
 \begin{equation}
 \Gamma_{1}(1 -  { e}^{-\mu} ) = { e}^{\mu} -1 -  U(\mu)
 =  { e}^{\mu} + 1 - r\lambda \,. \label{cond:Gamma1}
\end{equation}
Taking $\Gamma_1$ from \eqref{sol:Gammam} and inserting it
into (\ref{cond:Gamma1}) we obtain 
\begin{equation}
\label{rlm}
     r\lambda +  ({ e}^{\mu} - 1)
 \frac{ J_{r+1} (\Lambda)} {J_{ r} (\Lambda)} =  { e}^{\mu} + 1 \,.
\end{equation}
Using $U=r\lambda-2$ and extracting $r\lambda$ from \eqref{rlm} we arrive at the announced implicit equation (\ref{sol:Umax}) for $U(\mu)$.

\section{The case of a driven polymer}
  \label{App:asym}

Here we  generalize the analysis of  Section~\ref{sec:discrete} to the case of an asymmetric motion of the translocating polymer.  We  suppose
that the polymer is driven by a non-zero force $F$  through the pore: 
  this can be modeled by saying  that the polymer hops  to the right
 with rate $p$ and  to the left with rate $q$ with $p/q = e^F$.
 The basic equations now become
\begin{eqnarray}
 \frac{d Q_j}{dt} &=&  p Q_{j-1} + q  Q_{j+1} + \lambda \sum_{k>j} Q_k
 -\left[p+q+\lambda(j-1) \right]  Q_j 
 \, \hbox{ for } \,\,\, j>1, 
 \label{eq:Qjbis} \\
  \frac{d Q_1}{dt} &=& q  Q_2 +  \lambda \sum_{k>1} Q_k - p Q_1 \, .
  \label{eq:Q1bis}
\end{eqnarray}
One  has to define $Q_0$ such that  $p Q_0 =  q Q_1$.  

In the  stationary state,
the equations for the  $E_m$'s defined as  $E_m = \sum_{k>m} Q_k $ become
\begin{eqnarray}
    p E_{m-1} +  q  E_{m+1} =  \left(p + q + \lambda m  \right) E_m \, , 
\end{eqnarray}
with $E_0 =1$ and   $p E_{-1} +  q  E_{1} =  p + q$. 
The solution of this difference equation that satisfies the boundary condition is
\begin{equation}
    E_m = \left( \frac{p}{q} \right)^{m/2}
 \frac{ J_{m+\Lambda}(\overline{\Lambda})}{J_{\Lambda}(\overline{\Lambda})} 
 \,\,\, \hbox{ with } \,\,\,  \Lambda =\frac{p+q}{\lambda} \quad\text{and}\quad
\overline{\Lambda}=\frac{2\sqrt{pq}}{p+q}\,\Lambda \, . 
\label{eq:solEmasym}
\end{equation}
This leads us to the velocity of the polymer:
\begin{equation}
 v = p +q( Q_1 -1) = p -qE_1 =
  p - \sqrt{pq}\,\, 
 \frac{ J_{\Lambda+1}(\overline{\Lambda})}{ J_{\Lambda}(\overline{\Lambda})}\,,  \quad 
\label{eq:vitessesym}
\end{equation}
 Similarly,  we derive 
 the following  equation for the maximal  eigenvalue and the corresponding
 eigenvector for the cumulant generating deformed matrix:
\begin{equation}
 p  { e}^{\mu}\Gamma_{m-1} +  q { e}^{-\mu}
\Gamma_{m+1}  = 
  \left(p +q + \lambda m +  U(\mu) \right)\Gamma_m 
  \, ,  
\label{eq:Gammaasym}
\end{equation}
which is  valid for $m \ge 0$  with boundary conditions $\Gamma_0 =1$
and $p (\Gamma_{-1} -1 ) + q { e}^{-\mu} (\Gamma_{1} -1 ) =0$. This leads us to the implicit equation for $U(\mu)$:
 \begin{equation}
U(\mu) =  ({ e}^{\mu} - 1)
 \left( p  -  \sqrt{pq}\,\, \frac{ J_{\Lambda(  1 + \frac{U}{p+q})   +1}(\overline{\Lambda})}
 {J_{ \Lambda(  1 + \frac{U}{p+q})}(\overline{\Lambda})} \right)  \, . 
\label{sol:Umaxasym}
\end{equation}
Expanding \eqref{sol:Umaxasym} to the first order in $\mu$ we recover velocity and 
an expansion to the second  order gives the diffusion constant:
\begin{equation}
  {\mathcal D} = \frac{v}{2} -  \sqrt{pq}\,\,\frac{v}{\lambda}
  \left( \frac{\partial}{\partial \Lambda}
  \frac{ J_{\Lambda+1}(x)} {J_{ \Lambda}(x)} \right)
 \Big|_{x =\overline{\Lambda}}
\label{formule:Dasym}
\end{equation}

\section{Asymptotic expansions for the continuous polymer model}
\subsection{Large $x$ behavior of $E(x)$}
 \label{app:largex}

To compute the large $x$ asymptotic, we  apply  the steepest decent technique 
 on Eqs.~\eqref{ansatz}--\eqref{Zu-sol}. These equations  show that 
we need to integrate $e^{F(u)}$ along the contour in the complex $u$ plane with
\begin{equation}
\label{Fu}
F(u) = u(x-1) - \frac{u^3}{3\lambda} + \int_0^u dv\,\frac{e^v-1}{v}
\end{equation}
We must find the stationary point $u_*$ where $F'(u_*)$ vanishes and deform 
the contour so that it  passes 
 through  the stationary point in the direction of steepest decent. 
Using \eqref{Fu} we get
\begin{equation}
\label{u*}
x-1-\frac{u_*^2}{\lambda}+\frac{e^{u_*}-1}{{u_*}}=0
\end{equation}
for the critical point and we see that the direction of the steepest decent
 is along the imaginary axis. When $x\gg 1$, we have 
\begin{equation}
\label{u*-sol}
u_* = -\sqrt{\lambda(x-1)}-\frac{1}{2(x-1)}+\mathcal{O}(x^{-5/2})
\end{equation}
Writing $u=u_*+iw$ and knowing  that the constant that appears in
\eqref{Zu-sol} is equal to $(-iA/2),$ we recast \eqref{ansatz}--\eqref{Zu-sol} into 
\begin{equation}
\label{Gauss}
E=\frac{A}{2}\,e^{F_*}\int_{-\infty}^\infty dw\,
\exp\!\left(-\sqrt{\frac{x-1}{\lambda}}\,w^2\right)
\end{equation}
where $F_*=F(u_*)$. Computing the Gaussian integral in \eqref{Gauss}
 and using \eqref{u*-sol} to simplify $F_*$ we deduce 
\begin{equation*}
E=A\,\frac{e^{\gamma_E} \sqrt{\pi}}{2}\,\lambda^{3/4}(x-1)^{1/4}
\exp\!\left[-\frac{2}{3}\,\sqrt{\lambda}\,(x-1)^{3/2}\right]
\end{equation*}
where $\gamma_E=0.5772\ldots$ is the Euler's constant. 

\subsection{Large $\lambda$ behavior of the velocity}
 \label{App:Largelambda}

We calculate the leading behavior of the velocity in the $\lambda \to \infty$ limit
by studying separately the numerator and the denominator in  Eq.~\eqref{eq:Formulev}. Keeping only  the dominant terms in   $\lambda$, we find, using  Eqs.~\eqref{def:W1}--\eqref{def:W2}, 
\begin{eqnarray}
 \int_0^\infty dw\,\mathcal{W}_1(w)\,C(w) &\simeq& \lambda  \int_0^\infty dw\, \frac{C(w)}{w^3}
 \left(\sin(w+W)-\left(1-\frac{w^2}{2}\right)\sin W-w\cos W \right) \,\, +  \, {\mathcal O}(1)  \, ,
 \label{domW1} \\
 \int_0^\infty dw\,\mathcal{W}_2(w)\,C(w) &\simeq&  -\lambda  \int_0^\infty dw\, \frac{C(w)}{w^2}
 \left( \cos(w+W) -\cos W +w\sin W \right)   \,\,  +  \,  {\mathcal O}(1)  \, . \label{domW2}
\end{eqnarray}
 The function $W(w)$, see \eqref{sol-C}, is  given by
 $$W(w) = \int_0^w dv\,\frac{\sin v}{v}  +  \frac{w^3}{3\lambda} \equiv W_0(w) +  \frac{w^3}{3\lambda} \, ,  $$
where we have denoted by  $W_0(w)$  the leading behavior. One could expect
that the  dominant behavior can be found by substituting  $W_0$ in place of $W$
 in the  expressions~\eqref{domW1} and~\eqref{domW2}. But in fact, we have
\begin{eqnarray}
 \int_0^\infty dw\, \frac{C(w)}{w^3} \left(\sin(w+W_0)-\big(1-\tfrac{w^2}{2}\big)\sin W_0-w\cos W_0 \right)
  &=& 0 \, ,  \label{domW1-0} \\
 \int_0^\infty dw\, \frac{C(w)}{w^2} \Big( \cos(w+W_0) -\cos W_0 +w\sin W_0 \Big)  &=& 0 \, .
  \label{domW2-0}
 \end{eqnarray}
We now explain  why these  integrals  vanish  identically. Because 
the integrands are even functions of $w$ (recall that $C(w)$ is even 
and  $W_0(w)$ is odd), we can replace the lower bound $0$  in both the integrals  by $-\infty$. 

\begin{figure}[ht]
\begin{center}
\includegraphics[width=0.25\textwidth,angle=-90]{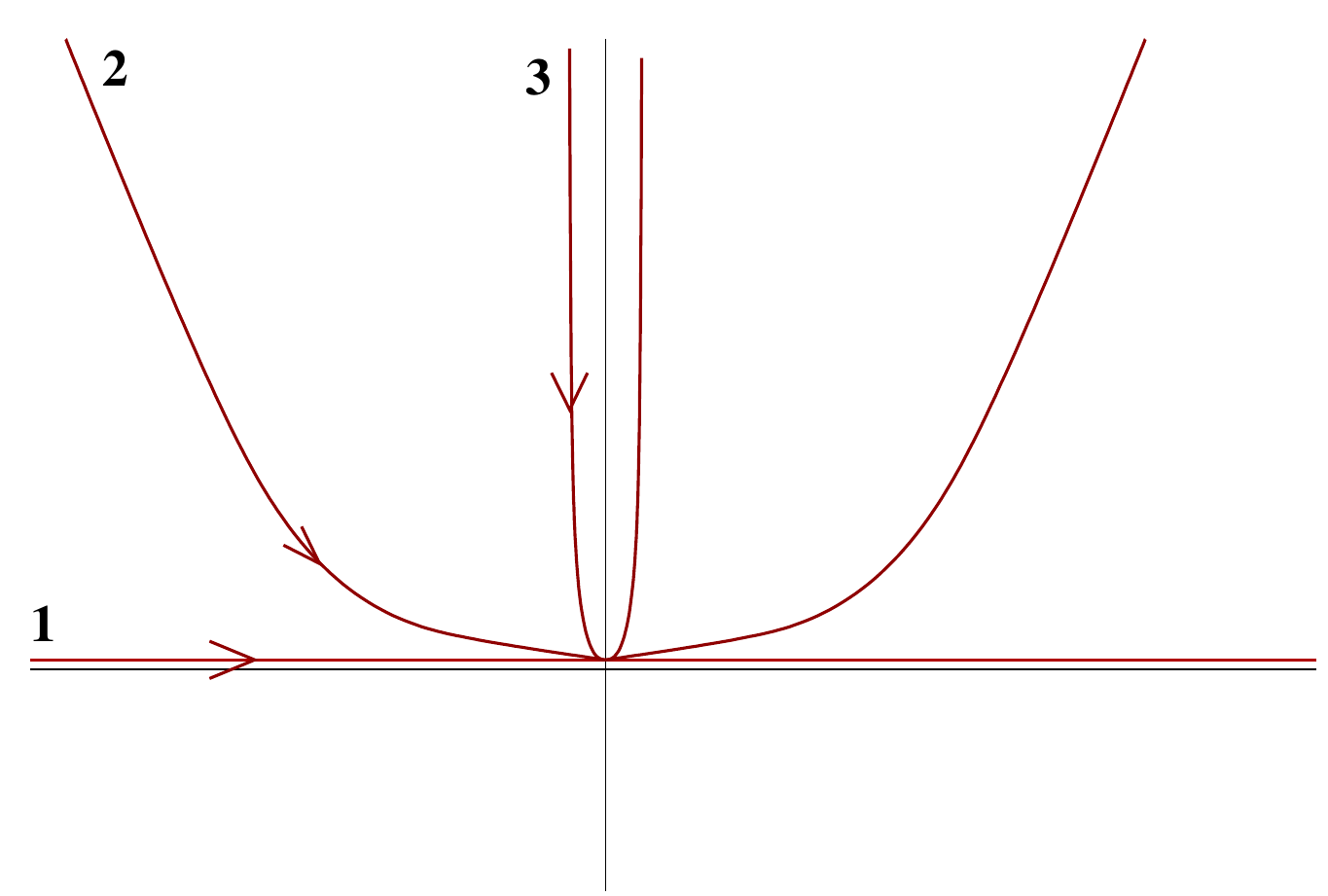}
\caption{Contour deformation in the upper complex plane used to prove  Eqs.~\eqref{domW1},~\eqref{domW1-0}
and \eqref{eq:auxil2}. The final contour follows the positive imaginary axis  down from $+i\infty$ to 0 and then back from 0 to  $+i\infty$. There are no singularities or cuts on the imaginary axis, therefore the two halves of the contour simply cancel with  each other.}
\label{Fig:Cdeformation}
\end{center}
\end{figure}

Before deriving \eqref{domW1-0}--\eqref{domW2-0}, we shall prove a much simpler identity. Let us show that the following relation is satisfied:
\begin{eqnarray}
  \int_0^\infty dw\, C(w)  \cos(w+W_0) =\frac{1}{2} 
 \int_{-\infty}^\infty dw\,  C(w)  \cos(w+W_0) = 0 \, .
  \label{eq:auxil}
 \end{eqnarray} 
We can  add  to this expression the following integral which vanishes because the integrand is odd
 \begin{eqnarray}
 \frac{i}{2}  \int_{-\infty}^\infty dw\,  C(w)  \sin(w+W_0) = 0 \, .
\end{eqnarray}
 Then, using    the explicit expressions \eqref{sol-C} for  $C(w)$  and  $W_0(w)$,
 and defining the function 
\begin{equation}
{\mathcal E}(w) = \int_0^w dv\,\frac{e^{i v} - 1}{v} \, ,
 \label{def:GothE}
 \end{equation}
the identity~\eqref{eq:auxil} becomes equivalent to
 \begin{eqnarray}
 \int_{-\infty}^\infty dw\,  C(w) e^{i(w+W_0)} = \int_{-\infty}^\infty dw\, 
   e^{{\mathcal E}(w)}  e^{ i w}  = 0 \, .
 \label{eq:auxil2}
 \end{eqnarray}
This integral converges in the $w$-complex plane for $w$ with positive imaginary
value (i.e. $\Im(w) \ge 0$). The value of the integral will therefore be unchanged if we deform the real axis $(-\infty,+\infty)$ into any contour located in the upper half-plane. One can take for example $w = y + i b$ with $-\infty<y<\infty$ and a fixed, strictly positive, value of $b$. Substituting this expression for $w$ in Eq.~\eqref{eq:auxil2}, we observe that  $e^{-b}$ appears as a prefactor and that the rest of the  integral remains bounded as  $b$ varies; taking the $b \to +\infty$ limit proves that the value of   integral is indeed 0. A more elegant way to prove that the integral  in Eq.~\eqref{eq:auxil2}  vanishes  identically  is to deform
the real axis as shown in Fig.~\ref{Fig:Cdeformation}.

Utilizing the same approach, one can prove the following identities:
\begin{eqnarray}
  \int_{-\infty}^\infty dw\,   e^{{\mathcal E}(w)}  \,   \frac{e^{i w} -1}{w}  &=&  0 \, ,  \\
    \int_{-\infty}^\infty dw\,   e^{{\mathcal E}(w)} \,  \frac{e^{i w} -1 -iw}{w^2}  
    &=&  0 \, ,  \label{eq:id2ndorder}\\
    \int_{-\infty}^\infty dw\,   e^{{\mathcal E}(w)} \,  \frac{e^{i w} -1 -iw +w^2/2}{w^3}  &=&  0 \, .
 \label{eq:id3rdorder}
\end{eqnarray}
Note that  Eq.~\eqref{domW2-0} is the  real  part of   
Eq.~\eqref{eq:id2ndorder} and  Eq.~\eqref{domW1-0} is the imaginary  part of   Eq.~\eqref{eq:id3rdorder}.

In order to find the asymptotic expression for the velocity in the large $\lambda$ limit, it is thus not enough to replace $W$ by $W_0$ in the  expressions~\eqref{domW1} and~\eqref{domW2} and we must  perform an expansion to the next order. The leading behavior of the integral  $\int dw\,\mathcal{W}_1(w)\,C(w)$ is obtained by subtracting from the right-hand side of Eq.~\eqref{domW1} the left-hand side of Eq.~\eqref{domW1-0}. There are three terms which we now analyze separately. For the first term we obtain 
\begin{eqnarray*}
 \int_0^\infty dw\, \frac{C(w)}{w^3}
 \left(\sin(w+W)-  \sin(w+W_0) \right) &=&  2  \int_0^\infty dw\, \frac{C(w)}{w^3} \, 
 \sin\!\left(\frac{w^3}{6\lambda}\right)
  \, \cos\!\left(w + \frac{w^3}{6\lambda}+W_0\right)  \\
   &=& \frac{2}{\lambda^{2/3}} \int_0^\infty dx \, \frac{C(\lambda^{1/3}x)}{x^3} \sin\frac{x^3}{6}
  \cos\!\left(\lambda^{1/3}x+ \frac{x^3}{6}+W_0(\lambda^{1/3}x)\right)
 \end{eqnarray*}
where in the second line we have made the change of variables $w \to \lambda^{1/3} x$.  For large values of the argument $w$, we have $C(w) \simeq  e^{\gamma_E}/w$  and $W_0(w) \to \pi/2$. Therefore, the above term is of order 
${\mathcal O}(\lambda^{-1})$. 
  
For the second term, we find 
 \begin{eqnarray*}
      -\int_0^\infty dw\, \frac{C(w)}{w^3}( 1 -\frac{w^2}{2}) \left(\sin(W)-  \sin(W_0) \right)
  &=&  \int_0^\infty dx \,     \frac{C(\lambda^{1/3}x)}{x^3} 
        \left(x^2 - \frac{2}{\lambda^{2/3}} \right)\sin\frac{x^3}{6}
        \cos\!\left( \frac{x^3}{6}+W_0 \right)   \\
 &\simeq& -\frac{e^{\gamma_E}}{\lambda^{1/3}} \int_0^\infty dx \, 
 \left( \frac{\sin x^3/6}{x} \right)^2 
 \label{termedominant}
 \end{eqnarray*}
where in the last equation we have inserted the asymptotics of 
$C$ and $W_0$ for large values of the argument and kept only the dominant term.  

Finally the third term reads
\begin{eqnarray*}
 \int_0^\infty dw\, \frac{C(w)}{w^2}\left(\cos W_0 - \cos W \right) =
 \frac{2}{\lambda^{1/3}} \int_0^\infty dx \, 
 \frac{C(\lambda^{1/3}x)}{x^2} \sin\frac{x^3}{6}
  \sin\!\left(\frac{x^3}{6}+W_0(\lambda^{1/3}x) \right)
  \end{eqnarray*}
and it scales as $\lambda^{-2/3}$ since $C(w) \sim w^{-1}$ for large $w$.
 
Comparing these asymptotics we see that the leading contribution arises from the second term.  Thus the integral \eqref{domW1} is equal to 
\begin{equation}
     \int_0^\infty dw\,\mathcal{W}_1(w)\,C(w) = -\lambda^{2/3} e^{\gamma_E}
  \int_0^\infty  dx \, \left( \frac{\sin x^3/6}{x} \right)^2  = 
 -\lambda^{2/3}\,\frac{e^{\gamma_E}\, 3^{1/6}\, \Gamma(2/3)}{4} 
 \label{LargelequivW1}
\end{equation}
in the leading order.  Using a similar reasoning, we find the leading behavior of the integral \eqref{domW2} 
\begin{equation}
     \int_0^\infty dw\,\mathcal{W}_2(w)\,C(w)  =
  2 \lambda^{2/3} e^{\gamma_E}
  \int_0^\infty  dx \, \left( \frac{\sin x^3/6}{x} \right)^2 
 \label{LargelequivW2}
\end{equation}
Substituting \eqref{LargelequivW1}--\eqref{LargelequivW2} into the formula~\eqref{eq:Formulev} for the velocity shows that  $v \to 2$ when $\lambda \to \infty$ 
 as written in Eq.~\eqref{limitspeedLlambda}.

 \section{Generating function of  the continuous polymer}
  \label{App:deformed-operator}
  
In this appendix, we investigate the time evolution of the probability density $P(L,x,t)$ in the continuum setting. By definition, $P(L,x,t) dL dx$ represents  the probability that the total length of the translocated polymer inside the cell  is between $L$ and $L+dL$ and that the distance of the leftmost adsorbed chaperone from the pore is
between   $x$ and  $x + dx.$ The probability density $P(L,x,t)$ evolves according to
\begin{equation}
\frac{\partial P}{\partial t} = \frac{\partial^2 P}{\partial x^2}
 + 2 \frac{\partial^2 P}{\partial x \partial L} + \frac{\partial^2 P}{\partial L^2}
   -\lambda(x-1)\Theta(x-1) P
+\lambda\int_{x+1}^\infty dy\,P(L,y,t)\,.
\label{PLt_govern}
\end{equation}
The Heaviside function $\Theta(x-1)$ ensures that this equation is valid for all values of  $x > 0$. If we study what happens in the vicinity of the $x=0$ boundary, we observe that the correct  boundary condition is given by
 \begin{equation}
   \frac{\partial P}{\partial x} + \frac{\partial P}{\partial L}= 0 
 \quad \text{for} \quad x=0, \,\,  L \ge 0 \,. 
\label{eq:BCP}
\end{equation}
(This  boundary condition can be obtained by writing the equation in a discrete form near $x=0$.) It is useful to consider the Laplace transform of $P(L,x)$ with respect to $L$
\begin{equation}
  \hat P(\mu,x,t) = \int_0^\infty dL\, { e}^{\mu L}  P(L,x,t) \, ,
\end{equation}
and to use the cumulative function
 \begin{equation}
 {\mathcal E}(\mu,x,t) =  \int_x^\infty  dy\, \hat P(\mu,y,t) \, .
\end{equation}
This function ${\mathcal E}$ satisfies the evolution equation 
\begin{equation}
 \frac{\partial {\mathcal E} }{\partial t} = 
 \frac{\partial^2  {\mathcal E} }{\partial x^2} -\lambda(x-1)\Theta(x-1)  {\mathcal E}     
-\lambda\int_{\max(1,x)}^{x+1} dy\, {\mathcal E}(\mu,y) -
  2 \mu  \frac{\partial  {\mathcal E} }{\partial x} +  \mu^2  {\mathcal E}
 \equiv   {\mathcal L}_\mu  {\mathcal E} \, ,
 \label{eq:evolEgoth}
\end{equation}
as it follows from \eqref{PLt_govern}. The boundary condition 
is deduced from Eq.~\eqref{eq:BCP} to yield 
 \begin{equation}
   \frac{\partial^2  {\mathcal E}}{\partial x^2}(\mu, 0) 
 -\mu  \frac{\partial  {\mathcal E}}{\partial x}(\mu,x) = 0 \,. 
\label{eq:BC}
\end{equation}
We observe that  equation~\eqref{eq:evolEgoth} is very similar to 
equations~\eqref{large_E}--\eqref{small_E} satisfied by the cumulative probability $E(x,t)$,  the only difference being  in the terms containing $\mu$ and $\mu^2$. The differential  operator  ${\mathcal L}_\mu$ that governs the evolution of  ${\mathcal E}$  is a deformation with parameter $\mu$ of the evolution operator  ${\mathcal L}_0$ for $E$. The long time limit  of   ${\mathcal E}$ will be given by the dominant eigenvector which corresponds  to  the largest eigenvalue $U(\mu)$  of ${\mathcal L}_\mu$. We denote $E_\mu(x)$ this dominant eigenvector. It satisfies 
\begin{equation}
\left(U(\mu) - \mu^2\right){E}_\mu = \frac{d^2  { E}_\mu }{d x^2} 
  -\lambda(x-1)\Theta(x-1)  { E}_\mu     
  -\lambda\int_{\max(1,x)}^{x+1} dy\, { E}_\mu(y) -
  2 \mu  \frac{d { E}_\mu }{d x} \, .
\label{eq:domE}
\end{equation}
The boundary condition~\eqref{eq:BC} becomes
\begin{equation}
  E_\mu''(0) = \mu  E_\mu'(0) \, ,
 \label{eq:BCdomE}
\end{equation}
where the prime denotes the derivative with respect to  $x$. 
We also normalize  $E_\mu(0) =1$, which is an overall constant  that
does  not affect  the final result. We emphasize that this situation is perfectly analogous to the discrete case where the Markov operator  $\mathbb{M}(0)$   was $\mu$-deformed into  $\mathbb{M}(\mu)$ [see Eq.~\eqref{eq:defMmu}] and where the dominant eigenvalue allowed us to determine the cumulant generating function.

For $x \ge 1$, we obtain,  using again the Laplace method \cite{landau,goursat}, 
an integral representation for $E_\mu(x)$, which assumes a form analogous
to the one found in Eq.~\eqref{E-sol}
\begin{equation}
\label{Emu-sol}
\begin{split}
&E_\mu(x)    = A_\mu\int_0^\infty dw\,C_\mu(w)\,C_\mu(w,x) \\
&C_\mu(w)   =  \exp\!\left(\int_0^w dv\,\frac{\cos v-1}{v} \,\, 
                         -\frac{\mu}{\lambda}\,w^2\right) \\
&C_\mu(w,x) = \cos[w(x-1)+W_\mu]  \\
&W_\mu(w)   = W(w) + \frac{U(\mu) - \mu^2}{\lambda} w =
                         \frac{w^3}{3\lambda} + \int_0^w dv\,\frac{\sin v}{v} 
                          + \frac{U(\mu) - \mu^2}{\lambda} w \, . 
\end{split}
\end{equation}
We shall write $W_\mu$ instead of $W_\mu(w)$ in  equations below.
Note that for $\mu =0$ we recover the expressions in  Eq.~\eqref{sol-C}.
The constant $A_\mu$  is yet to be   determined. 

In the $0 \le x \le 1$ region, the solution is given by
\begin{equation}
\label{solEmupetit}
E_\mu(x) =  \frac{{ e}^{r_{+} x} + { e}^{r_{-} x}}{2}
 -B_\mu  \frac{{ e}^{r_{+} x} - { e}^{r_{-} x}}{r_{+} - r_{-}} 
 - \lambda A_\mu\int_0^\infty dw\,C_\mu(w)\Phi_\mu(w,x) 
\end{equation}
with $r_{\pm} \equiv r_{\pm}(\mu) = \mu \pm \sqrt{U(\mu)}$ and 
\begin{eqnarray}
\Phi_\mu(w,x) &=&  
\frac{w^2 - r_{+}r_{-}}{w(r_{+}^2 + w^2)(r_{-}^2 + w^2)}\,\sin(wx+W_\mu)
- \frac{2\mu}{(r_{+}^2 + w^2)(r_{-}^2 + w^2)}\,\cos(wx+W_\mu) \\
               &+& \frac{\sin W_\mu}{w(r_{+} - r_{-})}
\left[\frac{w^2{ e}^{r_{+}x}}{r_{+}(r_{+}^2 + w^2)} -  \frac{w^2{ e}^{r_{-}x}}{r_{-}(r_{-}^2 + w^2)}
 +\frac{1}{r_{-}} - \frac{1}{r_{+}}\right] \nonumber \\
      &-&  \frac{\cos W_\mu} { r_{+} - r_{-} }
 \left(\frac{{ e}^{r_{+}x}}{r_{+}^2 + w^2}- \frac{{ e}^{r_{-}x}}{r_{-}^2 + w^2}\right) \nonumber
\label{defPhimu}
\end{eqnarray}
Note again  the similarity of structure between these equations and 
\eqref{E-small-sol}--\eqref{def:phi}.
 
The constants  $A_\mu$  and $B_\mu$ can be determined by taking into account that $E_\mu(x)$ and its derivative are  continuous at $x=1$. These constraints lead to 
\begin{equation}
 \label{eq:AmuBmu}
\begin{split}
 A_\mu J_{1,\mu} &= \frac{{ e}^{r_{+}} + { e}^{r_{-}}}{2}
 - B_\mu  \frac{{ e}^{r_{+}} - { e}^{r_{-}}}{r_{+} - r_{-}} 
 \\
 A_\mu J_{2,\mu} &= B_\mu  \frac{r_{+}{ e}^{r_{+}} - r_{-}{ e}^{r_{-}}}{r_{+} - r_{-}} 
 - \frac{r_{+}{ e}^{r_{+}} + r_{-}{ e}^{r_{-}}}{2} 
 \end{split}
\end{equation}
where we have used shorthand notation 
\begin{equation*}
J_{p,\mu} = \int_0^\infty dw\,\mathcal{W}_{p,\mu}(w)\,C_\mu (w), \quad p=1, 2\\
\end{equation*}
and 
\begin{equation*}
\begin{split}
\mathcal{W}_{1,\mu}(w) &=   \cos W_\mu(w)   + \lambda\, \Phi_\mu(w,1)        \\
\mathcal{W}_{2,\mu}(w) &=   w\,\sin  W_\mu(w) - \lambda\, \frac{\partial \Phi_\mu(w,x) }{\partial x}\Big|_{x=1}
\end{split}
\end{equation*}
Equations~\eqref{Emu-sol}--\eqref{eq:AmuBmu} determine the eigenfunction $E_\mu(x)$ in terms of  the eigenvalue $U(\mu)$. For $\mu =0$, $U(0)=0$ and  it can be verified that   $E_\mu(x)$ is  identical to  $E(x)$ given by 
Eqs.~\eqref{AB1}--\eqref{def:W2}.

The  maximal  eigenvalue $U(\mu)$ is  implicitly  determined  by  the self-consistent
equation obtained by  substituting  $x=0$ into Eq.~\eqref{eq:domE},  imposing the boundary condition~\eqref{eq:BCdomE}, and using the normalization 
$E_\mu(0) =1$. One gets
\begin{equation}
\label{UmE}
   U(\mu) = \mu^2 -  \mu  E_\mu'(0) \, . 
\end{equation}
Using relation $\frac{\partial \Phi_\mu(w,x) }{\partial x}\Big|_{x=0} =0$ [which can be deduced from \eqref{defPhimu}] and specializing Eq.~\eqref{solEmupetit} 
to $x=0$ we find $E_\mu'(0) = \mu - B_{\mu}$. Inserting this into \eqref{UmE} we simplify it to 
\begin{equation}
\label{UmB}
 U(\mu) =  \mu  B_{\mu} \, .
\end{equation}
Solving \eqref{eq:AmuBmu} and comparing with \eqref{UmB} we arrive at an implicit expression for the dominant eigenvalue  $U(\mu)$,
\begin{equation}
 U(\mu) =  \mu \left( \frac{r_{+} -r_{-}}{2} \right) 
 \frac{ \left(r_{+}{ e}^{r_{+}} + r_{-}{ e}^{r_{-}}\right) J_{1,\mu}
 +  \left( { e}^{r_{+}} + { e}^{r_{-}} \right)  J_{2,\mu}}
{  \left( r_{+}{ e}^{r_{+}} - r_{-}{ e}^{r_{-}}\right) J_{1,\mu}
 +   \left({ e}^{r_{+}} - { e}^{r_{-}}\right) J_{2,\mu}} \, ,
\nonumber 
\end{equation}
which, using  that $r_{+} -r_{-} = 2 \sqrt{ U(\mu)},$ can be rewritten as 
\begin{equation}
 U(\mu) =  \left[\mu\,
 \frac{ \left(r_{+}{ e}^{r_{+}} + r_{-}{ e}^{r_{-}}\right) J_{1,\mu}
 +  \left( { e}^{r_{+}} + { e}^{r_{-}} \right)  J_{2,\mu}}
{  \left( r_{+}{ e}^{r_{+}} - r_{-}{ e}^{r_{-}}\right) J_{1,\mu}
 +   \left({ e}^{r_{+}} - { e}^{r_{-}}\right) J_{2,\mu}}\right]^2.
\label{eq:SolUmuimplicit}
\end{equation}

We know from Eq.~\eqref{dvpteigenval} that  $U(\mu)$ generates the cumulants  of $L$ in the long time limit: $U(\mu) =  \mu v +  \mu^2 {\mathcal D} + \ldots$. 
Due to the presence of the overall factor $\mu$ in Eq.~\eqref{eq:SolUmuimplicit}, we observe that the expansion of  $U(\mu)$ with respect to  $\mu$ can be generated order by order in a self-consistent manner. At lowest order, taking the limit $\mu =0$ on the right-hand side of Eq.~\eqref{eq:SolUmuimplicit}, we retrieve the formula~\eqref{eq:Formulev}  for the velocity $v$. 
At the next order, retaining  the terms  linear in $\mu$  in $r_{+},r_{-}, \mathcal{W}_{1,\mu},\mathcal{W}_{2,\mu}$
and  $C_\mu $ (and substituting $U(\mu) \rightarrow  \mu v$) we  obtain a formula
for the diffusion constant ${\mathcal D}$:
\begin{equation}
{\mathcal D} = 1 - v + \frac{1}{3} v^3 + 
 \frac{ \Delta_2 -  v (\Delta_1+\Delta_2)}
{\int_0^\infty dw\,[\mathcal{W}_1(w)+\mathcal{W}_2(w)]\,C(w)}
 \label{eq:formuleDcont}
\end{equation}
with 
\begin{equation*}
\begin{split}
\Delta_1 & = \int_0^\infty dw\,\left[\mathcal{W}_1(w)\,\delta{C}(w) + 
                    \delta{\mathcal{W}_1}(w)\, {C}(w)\right]  \\
\Delta_2 & =  \int_0^\infty dw\,\left[\mathcal{W}_2(w)\,\delta{C}(w) + 
                     \delta{\mathcal{W}_2}(w)\, {C}(w)\right]
\end{split}
\end{equation*}
The functions $\delta{C}(w)$, $\delta\mathcal{W}_1(w)$ and 
$\delta\mathcal{W}_2(w)$ represent the first order derivatives  of the functions 
$C(w)$,  $\mathcal{W}_1(w)$  and  $\mathcal{W}_2(w)$ with respect to the parameter $\mu$. For instance, 
$\delta{C}(w)=\frac{\partial C_\mu(w)}{\partial \mu}\big|_{\mu=0}$ and its explicit expression reads
\begin{equation*}
 \delta{C}(w) =  -\frac{w^2}{\lambda} {C}(w).
\end{equation*}
The explicit expressions for $\delta\mathcal{W}_1(w)$ and 
$\delta\mathcal{W}_2(w)$ are quite cumbersome
\begin{eqnarray*}
\delta\mathcal{W}_1(w)  =  -  \frac{v}{\lambda} w \sin{W} 
 + \frac{\lambda}{w} \Big\{   - \frac{v}{w^4}  \sin(w+W)
   +    \sin{W}  \left(\frac{1}{3} +  \frac{v}{24} +\frac{v}{\lambda w} - \frac{2+v/2}{w^2}
  +  \frac{v}{w^4}    \right)
 \nonumber \\    + \cos(w+W) \left(  \frac{v}{\lambda w^2} -\frac{2}{w^3} \right) 
+   \cos{W}  \left(  \frac{v}{2\lambda}  - \frac{1 + v/6}{w} - \frac{v}{\lambda w^2}
 +  \frac{2+ v}{w^3} \right)
  \Big\}  \label{formuledeltaW1} \, , \\
\delta\mathcal{W}_2(w)  =  \frac{v}{\lambda} w^2 \cos{W}
  -\frac{\lambda}{w} \Big \{ \sin(w + W) \left(-\frac{v}{\lambda w} +\frac{2}{w^2} \right)
  +    \sin{W}  \left( 1 + \frac{v}{6} +\frac{v}{\lambda w}-  \frac{2+ v}{w^2} \right)  \nonumber \\
   - \frac{v}{w^3} \cos(w+W) +   
   \cos{W} \left(  \frac{v}{\lambda}- \frac{2+v/2}{w}  +\frac{v}{w^3}\right) \Big \}
 \label{formuledeltaW2} \, .
\end{eqnarray*}
These formulas can be used to establish the limiting  behavior of the diffusion constant when $\lambda$ is very small or very large [see Eqs~\eqref{limitdiffusionsmalllambda} and \eqref{limitdiffusionLlambda}].


\begin{thebibliography}{10}
\bibitem{Notewatson}  
    See \cite{watson}, section 8-42, equation~(7). 
    Note that the polynomials $B_m(x)$ that appear in
    that equation, and  in our equation~\eqref{AsymptExp}, have 
    no relation with Bernoulli polynomials which are also denoted $B_m(x)$.
\bibitem{Airy}
    It seems plausible that the difference-differential
     analog \eqref{large_stat}--\eqref{small_stat}
    of the Airy equation should play a role
    in various adsorption models.
\end{thebibliography}
\end{document}